\documentclass[12pt]{article}
\usepackage[utf8]{inputenc}
\usepackage[margin=0.8in]{geometry}
\usepackage{blindtext, xfrac}
\usepackage[english]{babel}
\usepackage[T1]{fontenc}
\usepackage{titling}
\usepackage{amsmath}
\usepackage{amssymb}
\usepackage{cite}
\usepackage{graphicx}
\usepackage{booktabs}
\usepackage{physics}
\usepackage{nameref}
\usepackage{bbold}
\usepackage{textgreek}
\usepackage{hyperref}
\usepackage{gensymb}
\usepackage{float}
\usepackage[usenames,dvipsnames]{color}
\usepackage{svg}
\usepackage{placeins}

\definecolor{greeny}{rgb}{0,0,1}

\newcommand*\diff{\mathop{}\!\mathrm{d}}
\newcommand*\Diff[1]{\mathop{}\!\mathrm{d^#1}}

\title{Perturbative Method for Mutual Information and Thermal Entropy of Scalar Quantum Fields}

\author{Joseph Bramante$^{1,2,3}$ and Andrew Buchanan$^{1,2}$
\\
$^1$\small Department of Physics, Engineering Physics, \& Astronomy \protect\\ 
\small Queen's University, Kingston, ON, K7L 2S8, Canada
\\
$^2$\small Arthur B. McDonald Canadian Astroparticle Physics Research Institute, \protect\\ 
\small Kingston, ON, K7L 3N6, Canada
\\
$^3$\small Perimeter Institute for Theoretical Physics, Waterloo, ON N2J 2W9, Canada}

\begin{document}
\maketitle

\begin{abstract}

A new approach is presented to compute entropy for massless scalar quantum fields. By perturbing a skewed correlation matrix composed of field operator correlation functions, the mutual information is obtained for disjoint spherical regions of size $r$ at separation $R$, including an expansion to all orders in $r/R$. This approach also permits a perturbative expansion for the thermal field entropy difference in the small temperature limit ($T \ll 1/r$).

\end{abstract}

\tableofcontents

\section{Introduction}
It has been appreciated for some time that the entropy of a black hole is not an extensive quantity. It is proportional to the black hole's surface area $A$, rather than its volume \cite{Bekenstein:1973ur,Hawking:1976de},
\begin{equation}
    S_{\text{BH}} = \frac{A}{4G},
\end{equation}
where $G$ is Newton's gravitational constant. This equation is considered to be one hint for developing a theory of quantum gravity. Indeed, the non-extensive nature of the black hole's entropy has been linked to a possible breakdown in local quantum field theory from gravitational effects \cite{Cohen:1998zx,Hsu:2004ri,Li:2004rb,Bramante:2019uub,Bramante:2019exc,Banks:2019arz,Blinov:2021fzl}. This and similar inquiries may benefit from further work on the entropy of quantum fields.

In \cite{Srednicki_1993}, Srednicki noted a connection between the black hole entropy and the Von Neumann entropy for a quantum field theory, defined as the expectation value of the negative logarithm of the density operator $\hat{\rho}$ of the state,
\begin{equation}
S_{QFT}    = -\langle \ln(\hat{\rho})\rangle_{\hat{\rho}} = -\Trace(\hat{\rho}\ln(\hat{\rho})).
\end{equation}
In short, the entropy of a massless scalar field on a sphere follows a similar area law as black hole entropy \cite{Srednicki_1993}. For a general review, see Casini and Huerta \cite{Casini_2009Review}.

Unfortunately, the quantum field theoretic entropy on a sphere has an ultraviolet divergence which complicates its study. For instance, Srednicki had to impose an ultraviolet cutoff by discretizing the sphere with a finite lattice. Rather than deal with this divergence directly, some authors choose to study a related quantity, the mutual information of two spatially separated regions \cite{swingle2010mutual,Wolf_2008,Katsinis_2020,Tajik_2023}. Given a bipartite state with density operator $\hat{\rho}_{AB}$ and reduced density operator $\hat{\rho}_{A}$ and $\hat{\rho}_{B}$, the mutual information can be defined in terms of entropy,
\begin{equation}
    I =S(\hat{\rho}_{A})+S(\hat{\rho}_{B})-S(\hat{\rho}_{AB}).
\end{equation}

In this work, we will focus on the mutual information between two spheres separated by a distance $R$. In a $d$ dimensional scalar field theory, this quantity has been shown to be follow an area law similar to entropy, proportional to the surface areas of each sphere for large $R$ \cite{Casini_2009Review},
\begin{equation}
    I(R,A_1,A_2)\sim C\frac{A_1A_2}{R^{2d-2}},
\end{equation}
where $A_1$ and $A_2$ are the surface areas of the spheres.

Shiba studied this case numerically for a massless field by approximating a quantum field theory with a finite number of harmonic oscillators and found estimates for the coefficient $C$ in two and three dimensions \cite{Shiba_2012}. Calabrese et al. and Cardy developed an analytic method to find these coefficients and found good agreement with Shiba's result \cite{Calabrese:2009ez,Calabrese:2010he,Cardy_2013}. This was later extended by Ag\'on and Faulkner \cite{Agon2015} as well as Chen et al.~to arbitrary dimensions \cite{Chen_2018},
\begin{equation}
    I\sim \frac{\sqrt{\pi}\Gamma(d)}{4^d\Gamma(d+\frac12)}\frac{r_1^{d-1}r_2^{d-1}}{R^{2d-2}},
\end{equation}
where $r_1$ and $r_2$ are the radii of the spheres and $\Gamma$ denotes the Gamma function. These analytical computations involved computing a related quantity: the R{\'{e}}nyi Mutual information,
\begin{equation}
        I_{a}(A,B)=S_{a}(\hat{\rho}_{A})+S_{a}(\hat{\rho}_{B})-S_{a}(\hat{\rho}_{AB}).
\end{equation}
Here $S_a$ is the R{\'{e}}nyi entropy, which converges to standard entropy as $a\to 1$,
\begin{equation}
    S_{a}(\hat{\rho})=\frac{1}{1-a}\log(\Trace(\hat{\rho}^a)).
\end{equation}
References \cite{Cardy_2013,Agon2015,Chen_2018} used a replica trick to compute $S_a$ for integer $a\geq 2$ and analytically continued to $a=1$ to compute the leading coefficient for mutual information. This is notably different than the strategy Shiba pursued using numerical calculations on a lattice. 
\\

The first goal of this paper is to present an alternative technique to exactly compute these leading coefficients and all following coefficients which is more in line with numerical lattice techniques, but does not rely on either analytical continuation or the replica trick. 
\\

This alternative technique will also allow us to compute the entropy difference for thermal fields. Parallel to the work on mutual information, there has considerable effort on understanding the divergences of the entropy on a sphere and the ways to regularize it \cite{Casini_2009Review}. The most straightforward way to do this is to simply subtract the entropy of an excited state on the sphere with that of the vacuum,
\begin{equation}
    \Delta S  = S(\hat\rho)-S(\hat\rho_0),
\end{equation}
where $\hat\rho$ is the  density operator of the state of interest and $\hat\rho_0$ is the density matrix of the vacuum. This is reasonable because only the change in entropy has physical meaning. In an influential paper \cite{Casini_2008}, Casini provided an argument for the finiteness of this quantity. Casini also pointed out the connection this quantity may have with the black hole entropy.  A slight reformulation of the area law for black hole entropy is the Bekenstein bound. The entropy of a physical state should be bounded from above by the entropy of a black hole of the same size and with mass equal to its energy. This entropy is a constant times its energy times a characteristic length,
\begin{equation}
    S \leq \lambda E R.
\end{equation}
Casini provided an interpretation of this bound in quantum field theory, using the modular Hamiltonian $\hat{\mathcal{H}}$ of the vacuum $\hat{\rho_0}$,
\begin{equation}\label{eq:modHamdfn}
    \ln{\hat{\rho_0}}=-\hat{\mathcal{H}}  -c\hat{I}.
\end{equation}
Here, $c$ is some constant representing the free energy of the vacuum. In place of entropy is the entropy difference $\Delta S$ and in place of $\lambda ER$ is what we will call Casini's $K$,
\begin{equation}\label{kInt}
    K = \langle\hat{\mathcal{H}}\rangle_{\hat{\rho}}-\langle\hat{\mathcal{H}}\rangle_{\hat{\rho}_0}.
\end{equation}
Casini noted that the difference $\Delta S-K$ was equal to the relative entropy $S(\hat{\rho}||\hat{\rho_0})$ which is always nonnegative, ensuring the bound $\Delta S\leq K$,
\begin{equation}
    S(\hat{\rho}||\hat{\rho_0}) = \langle \ln(\hat{\rho})-\ln(\hat{\rho_0})\rangle_{\hat{\rho}}.
\end{equation}
\\

The second goal of this paper is to study the entropy difference and Casini's $K$ for a massless scalar thermal field partially traced to a spherical region, which is a relatively simple example of a nontrivial excited state. Specifically, we will construct an expansion for the entropy difference for such a thermal field at low temperature.
\\

We will handle both of these problems using the same framework. We will use the fact that both the vacuum on two spheres and a thermal field on one sphere are Gaussian, which are particularly easy to understand states. Hence, we can organize Gaussian field correlations into a $2 \times 2$ correlation matrix to compute the entropy, in an approach related to that of Martin and Vennin \cite{Martin:2021xml}. Furthermore, since both of our goals have the scalar vacuum as a limiting case, this allows us to undertake a perturbative approach.  For the mutual information of spheres computation, when the distance between the two spheres is infinite, the spheres do not interact, becoming two copies of a vacuum on a single sphere. For the entropy difference of thermal fields computation, a thermal field approaches the vacuum as temperature goes to zero. The massless scalar vacuum on a sphere is conformally equivalent to the vacuum on a half plane and is well understood as a result \cite{Casini_2010}.  We use the term nearly conformal to refer to families of states with this property of limiting to conformal states. 
\\

Our central aim is to create a general perturbative strategy to compute expansions of information theoretic and thermodynamic quantities for nearly conformal field theories.
The general strategy will be to compute an expression for the series coefficients for a finite lattice, which is equivalent to computing a series expansion for a finite dimensional Gaussian state. We then take the continuum limit to find our coefficients. We will apply this strategy to our two goals. 
\\

The outline of this paper is as follows. Section \ref{sec:classicalLattice} presents preliminary definitions for configuration and Hilbert spaces in our setup, along with associated correlation functions and projections. Section \ref{perturblattice} gives some background on Gaussian states and provides a general description of our perturbative method, including a definition of the skewed correlation matrix. Section \ref{sec:contLim} develops the skewed correlation matrix for a massless quantum scalar field on a sphere. Section \ref{petuebSphere} gives the perturbative expansion away from a state that is the real scalar vacuum on a sphere. Sections \ref{MISect} and \ref{TempSect} work out the general series for the mutual information of distant spheres and the temperature of a thermal field, respectively. Section \ref{AreaSect} gives a brief argument showing why one should expect an area law to appear for the entropy difference in a scalar field in general. In Section \ref{sec:conclusion} we present general conclusions and future directions. Appendix \ref{sec:gaussLattice} details general mathematical features of Gaussian states in lattice field theories, including the modular Hamiltonian. Appendix \ref{app:math} lays out some mathematical background on symplectic matrices, the Takagi factorization, and spherical and cylindrical symmetries of fields on a sphere, which may be useful to reference at certain points in the derivation.

\section{Lattice Field Theory and the Configuration Space}
\label{sec:classicalLattice}
In this section, we summarize the approach we use to treat Gaussian states in scalar lattice field theory. We then give the general formula for a perturbative expansion for the entropy and mutual information of Gaussian states on a lattice. For a detailed derivation of this series, see Appendix
\ref{sec:gaussLattice}. Understanding the Gaussian states on the lattice will then allow us to take the continuum limit in Section \ref{sec:contLim}. As such, much of the results given in this section will be presented in a way to make taking the continuum limit as natural as possible.

\subsection{The Configuration Space}
 Any quantum Hilbert space is the quantization of some configuration space, with classical variables promoted to operators in the quantum space. For a real scalar field on a finite lattice of size $N$, the field operator $\hat{\phi}_j$ on the $j$-th lattice point corresponds to a classical variable $\phi_j$ describing the value of the field at that point. The configuration space is therefore itself a vector space, isomorphic to the vector space $\mathbb{R}^N$. 

Since the configuration space is a vector space, one might ask whether there are any natural linear transformations on this vector space and indeed, there are. Consider the two point correlation functions of some state $\hat{\rho}$, defined as
\begin{align}\label{FiniteCor}
X_{jk}&=2\langle \hat{\phi}_j\hat{\phi}_k\rangle, &  P_{jk}&=2\langle \hat{\pi}_j\hat{\pi}_k\rangle, & (V_{off})_{j,k}&=\langle\{\hat{\phi}_j,\hat{\pi}_k\}\rangle.
\end{align}
Here, $\hat{\phi}_j$ denotes the field operators, $ \hat{\pi}_j$ denotes their conjugate momentum operators and $\langle\cdot\rangle$ denotes the expectation value with respect to $\hat{\rho}$. We can then define matrices $X, P, V_{off}$ on the configuration space whose entries are the two point correlation functions. These matrices will be the central tool we use to analyze entropy in lattice field theory.

However, note that we are now dealing with linear operators in both the quantum Hilbert space and the configuration space. To concisely distinguish between the two, this work will refer to quantum operators as `operators' and use hats: $\hat{O}$. We will use no hats for operators on the configuration space and refer to them as `matrices', or `linear mappings on the configuration space'. This convention seems natural for now. But in the continuum limit, we will see that the configuration space is infinite dimensional and linear operators on infinite dimensional spaces are typically not called matrices. Regardless, we will still use this convention in the continuum limit for consistency.

\subsection{Configuration Spaces and Tensor Products in Lattice Field Theory}\label{ClassicalTP}
To describe mutual information in lattice field theory using the formalism of the configuration space, we must understand how the configuration space of a tensor product behaves in our setup. More precisely,
suppose we have a quantum Hilbert space $H$ with corresponding finite dimensional configuration space $C$ which decomposes into a tensor product $H_1\otimes H_2$. Suppose $H_1$ and $H_2$ have their own configuration vector spaces $C_1$ and $C_2$. Furthermore, suppose the field operators in $H_1$ and $H_2$ are field operators in $H$ as well. We must find how $C$, $C_1$ and $C_2$ are related.

Since the configuration space is the set of field operators, this means $C_1$ and $C_2$ are subspaces of $C$. Because they are finite dimensional, this means that they are closed subspaces as well. Because  $H = H_1\otimes H_2$, we can create an orthonormal basis in $H$ by constructing a complete set of commuting observables out of the field operators in $H_1$ and those in $H_2$. In particular, this means that any field operator can be written as a linear combination of some $H_1$ and $H_2$ field operators. Put in terms of the configuration space, this means that a basis for $C$ can be constructed from elements in $C_1$ and $C_2$. We can also show that $C_1$ and $C_2$ do not intersect except at the zero vector. If they did, the corresponding field and conjugate momentum operators would be operators in both $H_1$ and $H_2$. This is impossible because it would imply that a operator on $H_1$ does not commute with an operator on $H_2$. Putting the above facts together, we see that $C=C_1\oplus C_2$. Therefore, a tensor product in the quantum space corresponds to a direct sum in the configuration space.

\subsection{Block Matrices of a Matrix in Lattice Field Theory}\label{block}

In this paper, it will frequently be useful to divide a matrix into sub-matrices or \textit{blocks}, such as 
\begin{equation}
   A = \begin{bmatrix}
        A_{11}& A_{12}\\ A_{21}& A_{22}
    \end{bmatrix}.
\end{equation}To make the generalization to the continuum limit as simple as possible, we provide a basis independent way to define block matrices in the configuration space. 

Suppose the configuration space $C$ decomposes as a direct sum of closed subspaces $C= \bigoplus_{i=1}^nC_i$. This includes the special case of $C=C_1\oplus C_2$. We can define the \textit{projection onto $C_i$} as the matrix $P_i$ from $C$ to $C_i$ such that $P_{i}\vectorbold{v}$ equals $\vectorbold{v}$ if $\vectorbold{v}\in C_i$ and equals $0$ if $\vectorbold{v}\in C_j$ for any $j \neq i$. We can define \textit{the dual of the projection $\bar{P}_i$} as the matrix that naturally embeds vectors from $C_i$ into $C$. Given a matrix $A$ and a decomposition of $C$, we define the \textit{block of $A$ from $C_i$ to $C_j$ with respect to that decomposition} as $A_{C_i,C_j} = P_{j}A\bar{P}_{i}$. 

In addition, we will often be considering matrices on the vector space $C^2=C\oplus C$ in this paper. This vector space can be decomposed as $C^2= \bigoplus_{i=1}^nC_i^2$, where $C_i^2=C_i\oplus C_i$. Therefore we can define, analogously to the above, blocks from $C_i^2$ to $C_j^2$.

This definition of a block matrix will be necessary to write the mutual information in a way that generalizes naturally to the continuum limit. As we will see, it will also be necessary to properly account for spherical and cylindrical symmetry in the continuum limit. But before we take the continuum limit, we must complete our description of the lattice field theory case in the next section.

\section{Perturbative Series for Entropy of Gaussian States in Lattice Field Theory}\label{perturblattice}
In this section, we summarize the results of Appendix \ref{sec:gaussLattice}, which gives an overview of Gaussian states. Gaussian states are a special subset of states which are completely defined by their two point correlation matrices, as well as their field expectation values. However, we will always assume that the latter are zero, meaning our Gaussian states are completely defined by only their two point correlation matrices. To keep things mathematically straightforward, we continue to focus on a scalar field on lattice of size $N$, giving us a configuration space isomorphic to $\mathbb{R}^N$. 

Similar to References \cite{Martin:2021xml,Martin:2021qkg,Adesso2010} will define a single matrix which we call the \textit{skewed correlation matrix}, which completely describes a Gaussian state and describe how it behaves under the operation of the partial trace. We then give general expressions for entropy and mutual information in lattice field theory in terms of the skewed correlation matrix. Lastly, we outline how to perturb the entropy of a state expressed in terms of the skewed correlation matrix, which is the central result we will use in the continuum limit. 

\subsection{The Skewed Correlation Matrix and the Partial Trace}\label{partialtrace}
We combine all of the correlation functions into one matrix which we call the skewed correlation matrix $\alpha$, written as
\begin{equation}\label{skewCorrMatMain}
        \alpha = \begin{bmatrix}
            V_{off}^T&P\\-X&-V_{off}
        \end{bmatrix}.
\end{equation}
Here the blocks are chosen to respect the decomposition $C^2=C\oplus C$. This matrix encodes all information about a Gaussian state and proves to be a natural way to describe the state. 

For example, suppose we have a tensor product space $H=H_1\otimes H_2$. As discussed in Section \ref{ClassicalTP}, we can write the configuration space as a direct sum $C=C_1\oplus C_2$. Then we can decompose $C^2=C_1^2\oplus C_2^2$. If we have a Gaussian state $\hat \rho$ on $H$ with skewed correlation matrix $\alpha$, then the partial trace of $\hat\rho$ onto $H_1$ is also Gaussian. In addition, the partial trace of $\hat\rho$ onto $H_1$ has skewed correlation matrix equal to the block of $\alpha$ from $C_1^2$ to $C_1^2$ with respect to the decomposition $C^2=C_1^2\oplus C_2^2$. 

We note that the eigenvalues of $\alpha$ are strictly imaginary, with absolute value greater than or equal to one.

\subsection{Entropy and Mutual Information}\label{sec:Ent/MI}
 In particular, if we define a function  $h(z)=\frac{z}{2}\arccot(z)+\frac{1}{4}\ln(-\frac{1}{4}(z^2+1))$, the entropy $S$ of a Gaussian state with skewed correlation matrix $\alpha$ is
\begin{equation}\label{EntGauss}
  S = \Trace_{C^2}\left[h(\alpha)\right],
\end{equation}
where $ \Trace_{C^2}$ is the trace over the finite dimensional space $C^2$.
 Suppose we split the skewed correlation matrix into blocks according to the decomposition described in Section \ref{block},
\begin{equation} 
    \alpha = \begin{bmatrix}
        \alpha_{11}& \alpha_{12}\\ \alpha_{21}& \alpha_{22}
    \end{bmatrix}.
\end{equation}
We can also write the the skewed correlation matrix $\alpha_0$ of the non-interacting state $\rho_1\otimes\rho_2$ as
\begin{equation} 
    \alpha_{NI} = \begin{bmatrix}
        \alpha_{11}& 0\\ 0& \alpha_{22}
    \end{bmatrix}.
\end{equation}
Then the mutual information of this state between the two systems is
\begin{equation}
     I(1,2)  =\Trace_{C^2}\left[h(\alpha_{NI})-h(\alpha)\right].
\end{equation}

Lastly, suppose we have a Gaussian state with skewed correlation matrix $\alpha$ along with a `vacuum' Gaussian state skewed correlation matrix $\alpha_0$ and modular Hamiltonian $\hat{\mathcal{H}}$. We can write the quantity $K$ which is defined in Equation \eqref{kInt} as

\begin{equation}\label{eq:CasiniK}
    K = \Trace_{C^2}\left[h'(\alpha_0)(\alpha-\alpha_0)\right].
\end{equation}

\subsection{The Perturbative Expansion for Entropy}\label{sec:perturbedLattcie}
Suppose that we have two Gaussian states with skewed correlation matrices $\alpha_0$ and $\alpha_1$\footnote{The following definitions and Eq.~\eqref{eq:perturbedNice} will apply to any two Gaussian states $\alpha_i,\alpha_j$. However, we are immediately labelling one of the states as the vacuum state, $\alpha_0$, since we will be most interested in perturbing the vacuum state.}. If we write the difference between the skewed correlation function as $\Delta\alpha = \alpha_1-\alpha_0$, we can write the entropy difference between the two states as a series:
\begin{equation}
\label{eq:perturbedNice}
    S_1-S_0=\Trace_{C^2}\left[h'(\alpha_0)\Delta\alpha\right]+\sum_{n=2}^\infty\frac{1}{2\pi n i }\oint_\xi \diff z \text{ }h'(z)\Trace_{C^2}\left[\left[(z-\alpha_0)^{-1}\Delta\alpha\right]^{n}\right],
\end{equation}
where $\xi$ is a contour that the subset of the imaginary line where $|z|>1$.
Note that, thanks to the appearance of the resolvent $(z-\alpha_0)^{-1}$, this series is only useful if the diagonalization of $\alpha_0$ or the resolvent itself is known. If this is the case, this formula can be used to calculate a perturbation series for the entropy $S_1$ for  small $\Delta\alpha$. Furthermore, as we will see, it is in a form that extends very naturally to the continuum limit.

\subsection{Casini's $K$ and Relative Entropy
}\label{popout} 
Consider the first term in Equation \eqref{eq:perturbedNice}, and let us specify that $\alpha_0$ is a vacuum state. Then this is precisely the value of Casini's $K$ given in Equation \eqref{eq:CasiniK}. This provides a nice demonstration of the usefulness of this quantity. Furthermore, if we recall the identity $\Delta S= K - S(1||0)$ where $S(1||0)$ is the relative entropy between the two states, we see that the remaining terms in our expansion make up the relative entropy. Therefore, we see that we have a useful expression to study the entropy difference, and moreover, we have obtained a series expansion for relative entropy,

\begin{equation}
\label{eq:peturbRelEnt}
    S(1||0 )=-\sum_{n=2}^\infty\frac{1}{2\pi n i }\oint_\xi \diff z \text{ }h'(z)\Trace_{C^2}\left[\left[(z-\alpha_0)^{-1}\Delta\alpha\right]^{n}\right].
\end{equation}

\subsection{Eigenbasis of Skewed Correlation Matrix via Modular Hamiltonian}\label{sec:modHamLattcieMain}
To use the peturbative series described in Equation \eqref{eq:perturbedNice} in practice, we need to to have some information about the eigenvalues of $\alpha_0$. Here, we summarize the results of Section \ref{sec:modHamLatt} which provides a strategy to understand the spectral decomposition of $\alpha_0$ which generalizes nicely to the case of a vacuum in scalar field theory.

Suppose we have some Gaussian state with skewed correlation matrix $\alpha_0$. Recall the definition of the modular Hamiltonian,\begin{equation}
    \ln{\hat{\rho_0}}=-\hat{\mathcal{H}}  -c\hat{I}.
\end{equation}
The modular Hamiltonian can be written in terms of $\alpha_0$ as
\begin{equation}
     \hat{\mathcal{H}} =  \hat{\vectorbold{V}}^Th'(\alpha_0)J\hat{\vectorbold{V}}.
     \label{eq:modham1}
\end{equation}
Here $\hat{\vectorbold{V}}$ is the vector valued operator $(\hat{\boldsymbol{\phi}},\hat{\boldsymbol{\pi}})$ and $J$ is the matrix,
\begin{equation}\label{eq:Jmain}
    J =\begin{bmatrix}
        0& I\\-I&0\\
    \end{bmatrix},
\end{equation}
where $I$ is the identity matrix on $C$. Now, suppose we have solved the eigenvalue problem for $h'(\alpha_0)$:
\begin{equation}
    h'(\alpha_0)\begin{bmatrix}
        v_{k,1}\\ v_{k,2}
    \end{bmatrix} = i\lambda_k \begin{bmatrix}
        v_{k,1}\\ v_{k,2}
    \end{bmatrix}.
\end{equation}
Here, we have separated the eigenvectors into its two components which are each an element of $C$. Then, given any matrix $A$ in $C^2$, we can write the entries in the $h'(\alpha_0)$ (and therefore $\alpha_0$) eigenbasis as
\begin{equation}\label{eq:alphaBasis}
    A_{k,k'} = \begin{bmatrix}
         v_{k',2}^\dagger&-v_{k',1}^\dagger  \end{bmatrix}A\begin{bmatrix}
        v_{k,1}\\  v_{k,2}
    \end{bmatrix}.
\end{equation}
This allows us to write the resolvent of $\alpha_0$ in the $\alpha_0$ eigenbasis as
\begin{equation}\label{eq:DiagonalResolvent}
    ((z-\alpha_0)^{-1})_{k,k'}= \frac{\delta_{k,k'}}{z+i\coth(\lambda_k)}.
\end{equation}
Solving the eigenvalue problem for $h'(\alpha_0)$ instead of $\alpha_0$ may seem strange for the case of lattice field theory, as both eigenvalue problems would be equally difficult. But, as we will see, in the continuum limit $h'(\alpha_0)$ has a more easily computed spectrum than $\alpha_0$ because the former is a differential operator and the latter is not. Therefore, understanding the spectrum of $h'(\alpha_0)$ will provide an indirect approach to understand the spectrum of $\alpha_0$.

\section{Scalar Field Theory as The Continuum Limit of Scalar Lattice Field Theory}
\label{sec:contLim}

This section addresses the definition of scalar quantum field theory in terms of the skewed correlation matrix introduced in the prior section, and proceeds to define the vacuum and finite temperature correlation matrices for scalar quantum field theory in Sections \ref{sec:Vaccuum} and \ref{thermal}. We then explicitly compute the Gaussian decomposition of the vacuum skewed correlation matrix introduced in Section \ref{sec:modHamLattcieMain}, and find it involves a Fourier-type transform. Following this decomposition, we will be able to perturbatively expand expressions for the entropy around the vacuum. The first three subsections deal with questions of mathematical rigor; readers interested only in the main treatment and correlation matrices could skip to Section \ref{sec:Vaccuum}.

\subsection{The Configuration Space for a Scalar Field}
\label{classVect}
For a scalar quantum field theory on an open set $\Omega$, the configuration space is also a vector space: the set of tempered distributions on $\Omega$. These can be informally defined as the set of generalized functions (e.g. Dirac delta function) that grow as fast as a polynomial at infinity. More formally, the set of tempered distributions $\mathcal{S}^*(\Omega,\mathbb{R})$  are the continuous linear functionals in the Schwartz space on $\Omega$, where the Schwartz space is written $\mathcal{S}(\Omega,\mathbb{R})$. The Schwartz space is the set of all infinitely differentiable functions $f$ from $\Omega$ to $\mathbb{R}$ such that multiplying any derivative of $f$ by any polynomial is bounded.

In the case of quantum field theory, the two point correlation functions for the field operators $\hat{\phi}$ and $\hat{\pi}$ are often informally defined as follows. 
\begin{align}\label{CorrFunsQFT}
    X(\vectorbold{x},\vectorbold{y})&=2\langle \hat{\phi}(\vectorbold{x})\hat{\phi}(\vectorbold{y})\rangle, &  P(\vectorbold{x},\vectorbold{y})&=2\langle  \hat{\pi}(\vectorbold{x})\hat{\pi}(\vectorbold{y})\rangle, &\hat{V}_{off}(\vectorbold{x},\vectorbold{y})&=\langle\{\hat{\phi}(\vectorbold{x}),\hat{\pi}(\vectorbold{y})\}\rangle.
\end{align}
These are not well-defined functions in the traditional sense, rather they are tempered distributions on $\Omega\times\Omega$, which is to say they are in the set of tempered distributions $\mathcal{S}^*(\Omega\times\Omega,\mathbb{R})$.

For the case of a lattice field theory, recall that it was useful to establish a correspondence between correlation functions and linear maps on the configuration space, which amounted to defining a matrix with entries equal to the correlation functions (this was the construction of the matrix $\alpha$). For a quantum field theory, we can do something analogous by defining a linear map with a two point correlation function as an \textit{integral kernel}. To do so, we can use the Schwartz kernel theorem, which guarantees a linear map $A:\mathcal{S}(\Omega,\mathbb{R})\to \mathcal{S}^*(\Omega,\mathbb{R})$ \cite{Schaefer_1986}.
The explicit form of this linear map can be expressed by writing the correlation functions as an integral kernel. For example, we can write the field correlation matrix $X$ as
\begin{equation}\label{kernel}
    X[\phi](\vectorbold{y}) = \int_\Omega \diff^d \vectorbold{x} \text{ } X(\vectorbold{x},\vectorbold{y})\phi(\vectorbold{x}).
\end{equation}
Recall that, for consistency with the lattice field theoretic case, we will still use the term `matrices' for these linear maps to distinguish them from quantum operators. In addition, although the domain of definition of these matrices is not the entire configuration space, we will still refer to them as linear maps on the configuration space for simplicity of terminology.

In addition, recall that if $C$ is the configuration space, in much of this work we are concerned with the skewed correlation matrix which is a linear map on $C^2=C\oplus C$. For quantum field theory, $C^2$ can be thought of as the set of two component generalized functions on $\Omega$ which have the general form:
\begin{equation}\label{vectorval}
    \phi(\vectorbold{x}) = \begin{bmatrix}
        \phi_1(\vectorbold{x})\\ \phi_2(\vectorbold{x})
    \end{bmatrix},
\end{equation}
where the components $\phi_1$ and $\phi_2$ are tempered distributions. As suggested by this representation of $C^2$, a matrix on $C^2$ can be represented as a $2\times 2$-matrix valued kernel $K(\vectorbold{y},\vectorbold{x})$ which operates on an element of $C^2$ analogously to Equation \eqref{kernel},
\begin{equation}\label{rep1}
    K(\vectorbold{x},\vectorbold{y}) = 
    \begin{bmatrix}
        K^{1,1}(\vectorbold{x},\vectorbold{y}) &  K^{1,2}(\vectorbold{x},\vectorbold{y}) \\K^{2,1}(\vectorbold{x},\vectorbold{y}) &  K^{2,2}(\vectorbold{x},\vectorbold{y}) \\
    \end{bmatrix}.
\end{equation}

\subsection{Tensor Products Spaces and Block Matrices for a Scalar Field }
\label{fieldBlocks}

Our discussion of tensor product spaces and their corresponding configuration spaces for lattices in Section \ref{ClassicalTP} is mathematically valid owing to the finite dimension of the lattice configuration space. However, we run into a problem if we try to generalize to quantum field theory. Namely, quantum field theories are only well defined on an open set.
If the open set $\Omega$ is the union of two open sets $\Omega_1$ and $\Omega_2$ that do not share a boundary, the finite dimensional case carries over exactly with $C$, $C_1$, and $C_2$ being the tempered distributions on $\Omega$, $\Omega_1$, and $\Omega_2$ respectively. In this case, we also have $C_1$ and $C_2$ are closed subspaces of $C$ such that $C=C_1\oplus C_2$. If the configuration space decomposes as a direct sum as described here, then the definition of a block matrix given in Section \ref{block} carries over to the case of quantum field theory. This will allow us to have a well defined notion of mutual information between fields on two sets that do not share a boundary.

Similarly, Appendix \ref{spher} describes how the configuration space when $\Omega$ is a sphere can be decomposed into spherical wave mode subspaces. As noted in the appendix, the definitions given in Section \ref{block} carry over in that case as well. 

However, suppose $\Omega_1$ and $\Omega_2$ are open sets that do share a boundary. Alternatively, suppose  $\Omega_2=\Omega-\Omega_1$ is not open. In the former case, we won't have $C=C_1\oplus C_2$.  In the latter case, there is no sensible way to define a quantum field theory on $\Omega_2$. Furthermore, in both cases we may not have $C_1$ be a subspace of $C$ at all. This prevent us from defining a notion of a block matrix perfectly analogously to the lattice field theoretic case. These messy technicalities are a large part of why the quantum field theoretic cases are treated here without full rigour. 

We will still need some notion of a block matrix however. For example, suppose that $\Omega$ is the whole space $\mathbb{R}^d$ and $\Omega_1$ is some open subset, such as a sphere or the union of two disjoint spheres. Furthermore, suppose that we know the skewed correlation matrix on the whole space. If we want to generalize the result of \ref{partialtrace} to a scalar field, we will still need some way to write a block matrix

To do so, we can \textit{formally} define the projector onto a region $\Omega_1$ by restricting the domain of its input,
\begin{equation}
   P_{1}[\phi]=\phi|_{\Omega_1}.
\end{equation}
We can write the dual of the projection, which acts on tempered distributions on functions defined on $\Omega_1$ formally as well,
\begin{equation}\label{Pbound}
   \bar{P}_{1}[\phi](\vectorbold{x})= \begin{cases}
        \phi(\vectorbold{x}) & \mathrm{if}\text{ } \vectorbold{x}\in\Omega_1\\
        0 & \mathrm{if}\text{ } \vectorbold{x}\in\Omega-\Omega_1
    \end{cases}.
\end{equation}
While we cannot comfortably define the other blocks of $A$, we can define the block matrix of $A$ from $C_1$ to $C_1$ as \begin{equation}
    A_{C_1,C_1} = P_{1}A\bar{P}_{1}.
    \label{eq:ac1c1}
\end{equation}

There is a problem with this definition however. We expect that every correlation matrix will take a vector in the Schwartz space to the set of tempered distributions. This is guaranteed by the Schwartz Kernel Theorem, as explained in Section \ref{classVect}. But, this is not the case for the momentum correlation matrix. To illustrate this, suppose $\Omega$ is the whole space and $\Omega_1$ is the unit sphere. As can be checked in Section \ref{sec:Vaccuum}, the momentum correlation matrix $P$ for the vacuum on the whole space satisfies the requirements of the Schwartz kernel theorem. Yet, its block matrix on the sphere, $P_{C_1,C_1}$, takes a constant function to a spherically symmetric function that asymptotes to $\sim \frac{1}{1-r}$ as it approaches the boundary of the sphere at $r=1$. Such a function cannot be a tempered distribution on $\Omega_1$.\footnote{When one writes a distribution as a function $f$, we mean the functional $s\to \int_\Omega d\vectorbold{x} \text{ }f s$. Since Schwartz functions are not in general zero at the boundary, this is almost always a divergent integral, meaning the functional is ill-defined.} 

The reason for this divergence at the boundary is the discontinuity at the boundary in Equation \eqref{Pbound}. This means that the image of $ \bar{P}_{1}$ is discontinuous and therefore not in $\mathcal{S}(\mathbb{R}^d,\mathbb{R})$ which is the domain of definition of $A$. Put another way, this means that the linear map $A_{C_1,C_1}$ (cf. Eq.~\eqref{eq:ac1c1}) is not well defined.

This divergence at the boundary is ultimately responsible for the divergence of thermodynamic quantities like the entropy on a sphere. Since we are computing quantities in quantum field theory known or expected to be finite, it is likely that a more rigorous treatment of the problems we tackle here would show that these divergences cancel out and do not affect the validity of the final result. Indeed, a quick review of following sections will show that the final calculations involve only well-defined linear maps on the configuration space, even though the intermediate steps may not. We will make this assumption going forward and write the block matrix formally for now.

Suppose that we have the skewed correlation matrix $\alpha$ for a Gaussian state on the whole space with components $X$, $P$, and $V_{off}$. We can now use our notion of the block matrix to write the skewed correlation matrix for an open subset $\Omega_1$ with configuration space $C_1$ as\begin{equation}\label{skewCorrMatMainBlock}
        \alpha = J\sigma = \begin{bmatrix}
            (V_{off}^T)_{C_1,C_1}&P_{C_1,C_1}\\-X_{C_1,C_1}&-(V_{off})_{C_1,C_1}
        \end{bmatrix}.
\end{equation}
In Sections \ref{sec:Vaccuum} and \ref{thermal}, we will apply this to the case of a scalar field in the vacuum and at finite temperature respectively, partial traced to a unit sphere.

\subsection{The Trace over the Configuration Space for a Scalar Field}\label{smooth}
For the case of lattice field theory, the meaning of a trace over the square of the configuration space $C^2=C\oplus C$ is unambiguous since the configuration space is finite dimensional. But for a quantum field theory on an open set $\Omega$, there is no well-defined notion of a trace in the space of tempered distributions in general. This is problematic because it makes it difficult to generalize the $\Trace_{C^2}$ appearing in Equation \eqref{eq:perturbedNice}. 

Despite this, suppose we have two skewed correlation matrices $\alpha_1$ and $\alpha_0$ on $C_\Omega^2$ and define $\Delta\alpha$ as the difference between them. We will justify a definition for the trace such that the trace appearing in Equation \eqref{eq:perturbedNice}, namely $\Trace_{C_{\Omega}^2}\left[\left[(z-\alpha_{0,\Omega,d})^{-1}\Delta\alpha\right]^{n}\right]$, is both well-defined and finite assuming $z\neq a i  $ for $|a|\geq1$. 

To ensure that $\Trace_{C_{\Omega}^2}\left[\left[(z-\alpha_{0,\Omega,d})^{-1}\Delta\alpha\right]^{n}\right]$ is well-defined and finite, three conditions are necessary. Firstly, the open set $\Omega$ must be bounded. Secondly, the kernel of $\Delta\alpha$ must have smooth entries on the closure of $\Omega$. The third condition is that $\alpha_0$ is densely defined in $L^2(\Omega)^2$ under the Hilbert space norm.

Here, $L^2(\Omega)$ is the Hilbert space of square integrable functions on $\Omega$ and $L^2(\Omega)^2=L^2(\Omega)\oplus L^2(\Omega)$. We also write $\mathcal{S}_\Omega=\mathcal{S}(\Omega,\mathbb{R})$ as the Schwartz functions on $\Omega$. We note that $\mathcal{S}_\Omega^2$ is a subset of $L^2(\Omega)^2$ and $L^2(\Omega)^2$ is itself a subset of $C_\Omega^2$. 

Since, $\alpha_0$ is densely defined in $L^2(\Omega)^2$, its resolvent is by definition a continuous endomorphism on $L^2(\Omega)^2$ if $z$ in not in the spectrum of $\alpha_0$. Since the spectrum of a skewed correlation matrix is always imaginary with modulus greater than or equal to one in lattice field theory, its spectrum should remain so in the continuum limit. Therefore, if $z\neq a i  $ for $a\geq1$ or $a\leq -1$, the resolvent should be bounded.

Next, consider the matrix $\Delta\alpha$ on $L^2(\Omega)^2$. Since $\Delta\alpha$ is smooth, has a smooth kernel, and is defined on a bounded set, we can conclude that $\Delta\alpha$ belongs to the set of \textit{trace class} operators \cite{DELGADO20211}. In particular, this means that it is a continuous endomorphism on $L^2(\Omega)^2$, its spectrum is countable, and its trace (i.e. the sum of its eigenvalues) is finite.

We can now pull in an important fact from functional analysis: the product of any trace class operator with a bounded operator is a trace class operator \cite{Simon_2014}.  Using the fact that $\Delta\alpha$ is trace class and $(z-\alpha_0)^{-1}$ is bounded, we see that the $n$-th power of $(z-\alpha_0)^{-1}\Delta\alpha$ is a trace class endomorphism in $L^2(\Omega)^2$ for any $n\geq1$.

But this is all for $L^2(\Omega)^2$, not the configuration space $C_\Omega^2$. In fact, we have still not addressed the problem that, in an abstract sense, there's not an obvious way to define a trace over a space of tempered distributions. So far we have only written this trace formally. But how can we actually define it?
Recall that $\Delta\alpha$, like any matrix on $C_\Omega^2$, is in reality a linear mapping from the two-component Schwartz space $\mathcal{S}_\Omega^2$ to $C_\Omega^2$. But, $\mathcal{S}_\Omega^2$ is a dense subspace of $L^2(\Omega)^2$ meaning its image under the continuous $L^2(\Omega)^2$-endomorphism $\Delta\alpha$ is a dense subspace of $L^2(\Omega)^2$. 

Therefore, by the continuity of $(z-\alpha_0)^{-1}$, we can conclude that both the preimage and the image of the matrix $[(z-\alpha_0)^{-1}\Delta\alpha]^n$ are dense subspaces of $L^2(\Omega)^2$. Therefore, there should exist some orthonormal basis of $L^2(\Omega)^2$ contained in both the image and the preimage. But, traces in Hilbert spaces are defined by using an orthonormal basis. This justifies defining the trace of the matrix to be its trace over $L^2(\Omega)^2$.

\subsection{The Correlation Matrix of the Vacuum}\label{sec:Vaccuum}
Consider a real massless scalar field $\hat\phi$ in $d+1$-dimensional  Minkowski space with Lagrangian density
\begin{equation}
    \mathcal{L} = \square \hat \phi,
\end{equation}
where $\square$ denotes the d'Alembertian operator.
We select a preferred time direction\footnote{Note the the choice of a preferred time direction implicitly selects a preferred Lorentz frame. All physical and geometric quantities such as temperature and distance will be measured in this Lorentz frame.} and construct the corresponding Hamiltonian as
\begin{equation}
   \hat H = \frac{1}{2}\int \Diff{d}\vectorbold{x}\left[ \hat{\pi}(\vectorbold{x})^2-\hat{\phi}(\vectorbold{x})\nabla^2\hat{\phi}(\vectorbold{x})\right].
\end{equation}
Generalizing from the case of finitely many oscillators given in Appendix \ref{genThermalOscillator}, this has the form of the Hamiltonian of finitely many coupled oscillators with the sum replaced by an integral and with $K$ being equal to the negative Laplacian $-\nabla^2$, which is indeed positive definite. 
Therefore, we extend the correlation functions of a generalized thermal coupled oscillator in Appendix \ref{genThermalOscillator} to a scalar field.
For the special case of the vacuum state, the nonzero blocks of correlation matrix are the following fractional powers of the negative Laplacian:
\begin{align}
    X =  (-\nabla^2)^{-\frac{1}{2}} && P = (-\nabla^2)^{\frac{1}{2}}.
\end{align}

However, these formulas for the correlation operators are inconvenient for actual computation.
For the vacuum case, explicit representations of these operators are found in the math literature on pseudo-differential operators \cite{Stinga2018UsersGT}. For $d>1$, we write these operators in terms of their action on a field $\phi$ as 
\begin{align}\label{vacCorrs}
    \begin{split}
    X_{0,d}[\phi](\vectorbold{y}) &=  \frac{\Gamma(\frac{d-1}{2})}{2\pi^{\frac{d+1}{2}}}\int_{\mathbb{R}^d}\Diff{d}{\vectorbold{x}}\frac{\phi(\vectorbold{x})}{|\vectorbold{x}-\vectorbold{y}|^{d-1}}
\\
    P_{0,d}[\phi](\vectorbold{y}) &= (d-1)\frac{\Gamma(\frac{d-1}{2})}{2\pi^{\frac{d+1}{2}}}\operatorname{p.\!v.}\int_{\mathbb{R}^d}\Diff{d} \vectorbold{x}\frac{\phi(\vectorbold{y})-\phi(\vectorbold{x})}{|\vectorbold{x}-\vectorbold{y}|^{d+1}}.
    \end{split},
\end{align}
where $\operatorname{p.\!v.}$ regulates integral divergences\footnote{The symbol $\operatorname{p.\!v.}$ represents the Cauchy Principal value which regularizes the divergence at the singularity at $\vectorbold{x}=\vectorbold{y}$. In this case, $\operatorname{p.\!v.}\int_{\mathbb{R}^d}$ means $\lim_{\epsilon\to 0}\int_{\mathbb{R}^d/B_{\epsilon}(\vectorbold{y})}$ where $B_{\epsilon}(\vectorbold{y})$ is the ball of radius $\epsilon$ and centre $\vectorbold{y}$}.
These formulae match the correlation functions found in the physics literature \cite{Peskin_Schroeder_1995}, which are computed using different methods than our approach of taking the continuum limit of lattice field theory.
Here, the subscripts $0,d$ on the operators label the temperature of the field and the dimensions respectively.

Now, suppose we partial trace this state onto the unit sphere $\Omega$. The corresponding correlation matrix is a block matrix of the full correlation matrix. Following the prescription of Section \ref{fieldBlocks}, the correlation matrix on $\Omega$ is simply the restriction of this operator to functions on $\Omega$, written as
\begin{align}
    \begin{split}
    X_{0,\Omega,d}[\phi](\vectorbold{y}) &=  \frac{\Gamma(\frac{d-1}{2})}{2\pi^{\frac{d+1}{2}}}\int_{\Omega}\Diff{d}{\vectorbold{x}}\frac{\phi(\vectorbold{x})}{|\vectorbold{x}-\vectorbold{y}|^{d-1}}
\\
    P_{0,\Omega,d}[\phi](\vectorbold{y}) &= (d-1)\frac{\Gamma(\frac{d-1}{2})}{2\pi^{\frac{d+1}{2}}}\left[\operatorname{p.\!v.}\int_{\Omega}\Diff{d} \vectorbold{x}\frac{\phi(\vectorbold{y})-\phi(\vectorbold{x})}{|\vectorbold{x}-\vectorbold{y}|^{d+1}}+\int_{\Omega^C}\Diff{d} \vectorbold{x}\frac{\phi(\vectorbold{y})}{|\vectorbold{x}-\vectorbold{y}|^{d+1}}\right],
\end{split}
\end{align}
where $\Omega^C$ is the complement of $\Omega$.
\subsection{The Correlation Matrix at Nonzero Temperature}\label{thermal}

Now, assume the entire field is thermalized at temperature $T$. 
Let $\alpha_{T,d}$ be the skewed correlation matrix for a $d$-dimensional field at temperature $T$ and let $\alpha_{T,\Omega,d}$ be the same for this state partial traced to the sphere. 
The kernels of the correlation matrices for the unit sphere are simply the kernels for the whole space restricted to the sphere. 
Once again generalizing from the case of a generalized thermal coupled set of oscillators given in Equation \eqref{GenThermHam} in Appendix \ref{sec:gaussLattice}, the correlation matrix for the whole space will have the form
\begin{align}
    X_T =  \frac{1}{\sqrt{-\nabla^2}}\coth(\frac{ \sqrt{-\nabla^2}}{2T}) && P_T =\sqrt{-\nabla^2}\coth(\frac{ \sqrt{-\nabla^2}}{2T}).
\end{align}
We know the correlation operators for the vacuum, which corresponds to $T=0$. This allows us to write the difference between the skewed correlation matrices $\Delta \alpha_{T,d}=\alpha_{T,d}-\alpha_{0,d}$ in terms of the negative Laplacian as
\begin{equation}
    \Delta \alpha_{T,d} =  \begin{bmatrix}
        0&\sqrt{-\nabla^2}(\coth(\frac{ \sqrt{-\nabla^2}}{2T})-1) 
   \\-\frac{1}{\sqrt{-\nabla^2}}(\coth(\frac{ \sqrt{-\nabla^2}}{2T})-1)  &0
    \end{bmatrix}.
\end{equation}
The $2\times 2$ matrix kernel of this $\Delta \alpha_{T,d}$ has a very simple representation in Fourier space:
\begin{equation}
     \hat{K}_{{d,T}}(\vectorbold{k_1},\vectorbold{k_2}) =  \begin{bmatrix}
        0&|\vectorbold{k_1}|(\coth(\frac{ |\vectorbold{k_1}|}{2T})-1) 
   \\-\frac{1}{|\vectorbold{k_1}|}(\coth(\frac{ |\vectorbold{k_1}|}{2T})-1)  &0
    \end{bmatrix}\delta(\vectorbold{k_1}-\vectorbold{k_2}).
\end{equation}
To write this kernel in position space, we need to use the convolution theorem and find the $d$-dimensional (inverse) Fourier transforms (in a distributional sense) of $f_{d,P}(\vectorbold{k_1})=|\vectorbold{k_1}|(\coth(\frac{ |\vectorbold{k_1}|}{2T})-1)$ and $f_{d,X}(\vectorbold{k_1})=\frac{1}{|\vectorbold{k_1}|}(\coth(\frac{ |\vectorbold{k_1}|}{2T})-1)$. 

For $d\geq 3$, these functions are absolutely integrable, meaning that the Fourier transform is well-defined in the sense of standard functions. Moreover, they have exponential decay at infinity implying their Fourier transforms are real analytic. For $d\leq 2$, however $f_{2,X}$ is not absolutely integrable and finding the corresponding kernel requires more care. For this reason and because we lose out on  useful properties of the kernels,\footnote{For example, Observing that $f_{2,P}=-\nabla^2f_{2,X}$ is infinitely differentiable in $2D$ and using unit analysis, one can conclude that $f_{2,X}(r)=AT\ln(Tr)+Tg(r)$ for some constants $A$, $B$ and smooth function $g$. This logarithm term is problematic. It is not smooth with a singularity at $r=0$. Furthermore, the $T\ln(T)$ makes general expansions much messier.} we will only consider thermal fields for $d\geq 3$. 

For $d=3$, these Fourier transforms can be easily calculated using a Hankel transform of order $\frac{1}2$. We write the nonzero blocks of $\alpha_{T,\Omega,3}$.
\begin{align}\label{thermd3}
    \begin{split}
    X_{T,\Omega,3}[\phi](\vectorbold{y}) &=  \frac{T}{2\pi} \int_{\Omega}\Diff{3} \vectorbold{x}\frac{\phi(\vectorbold{x})}{|\vectorbold{x}-\vectorbold{y}|\tanh(\pi T|\vectorbold{x}-\vectorbold{y}|)}
 \\
    P_{T,\Omega,3}[\phi](\vectorbold{y}) =&\operatorname{p.\!v.}\int_{\Omega}\Diff{3} \vectorbold{x}\left[\frac{\phi(\vectorbold{y})}{\pi^2|\vectorbold{x}-\vectorbold{y}|^4}-\frac{\pi T^3\phi(\vectorbold{x})}{|\vectorbold{x}-\vectorbold{y}|\tanh(\pi T|\vectorbold{x}-\vectorbold{y}|)\sinh^2(\pi T|\vectorbold{x}-\vectorbold{y}|)}\right]\\&+\int_{\Omega^C}\Diff{3} \vectorbold{x}\frac{\phi(\vectorbold{y})}{\pi^2|\vectorbold{x}-\vectorbold{y}|^4}
    \end{split}
\end{align}

\subsection{The Spectrum of the Vacuum Skewed Correlation Matrix via a Local Modular Hamiltonian}\label{sec:Diagsphere}
We wish to generalize the expansion in Section \ref{sec:perturbedLattcie} to the case of quantum field theory, where the unperturbed state is the vacuum on the sphere with skewed correlation matrix $\alpha_{0,\Omega,d}$. To do so, we must understand the spectral decomposition of $\alpha_{0,\Omega,d}$ so that we can construct its resolvent $(z-\alpha_{0,\Omega,d})^{-1}$. 
If the configuration space was finite dimensional, diagonalizing $\alpha$ would be straightforward. But, in a quantum field theory on an open set $\Omega$, it is much more difficult to diagonalize $\alpha$ directly if $\Omega$ is not the whole space $\mathbb{R}^d$. If $\Omega$ is a sphere, we can see from Section \ref{sec:Vaccuum} that $\alpha_{0,\Omega,d}$ is some mix of a pseudo-differential operator and an integral operator. The spectra of these operators is generally not well understood.\footnote{There may actually be a strategy to directly diagonalize $\alpha_{0,\Omega,d}$ by representing it in terms of matrix valued Wiener-Hopf operators \cite{Bottcher2002}. However, this only applies if the the other method we propose works anyway (i.e when the modular Hamiltonian is local) and would be much more difficult.} 

Fortunately, $\alpha_{0,\Omega,d}$  belongs to a special case which allows us to diagonalize it by generalizing the result of Section \ref{sec:modHamLattcieMain}. If $\Omega$ is a half space or if $\Omega$ is a sphere and the quantum field theory is conformally invariant, the modular Hamiltonian is local. The massless scalar field is such an example of a conformally invariant theory. In this case, the positive definite matrix appearing in the modular Hamiltonian $h'(\alpha_{0,\Omega,d})J$ is a differential operator, where $J$ is the operator defined in Equation \eqref{eq:Jmain} and $h'(t)=\frac{1}{2}\arccot{(t)}$. Therefore $h'(\alpha_{0,\Omega,d})$ is also a differential operator. So, in principle, the eigenvalue problem of $h'(\alpha_{0,\Omega,d})$ should be equivalent to but much easier to consider than that of $\alpha_{0,\Omega,d}$ directly.

Casini and Huerta studied the entropy of an $n$-sphere in a scalar vacuum and among their work they give the explicit formula for the modular Hamiltonian of this state \cite{Casini_2010}.  Comparing Equations 7 and 8 in Casini and Huerta's work\footnote{There actually is small error in Casini and Huerta's result: a missing boundary term in Equation 7. We did not exposit on it here for space and because the missing term does not affect our final result.} with our formula for the Modular Hamiltonian in Equation \eqref{eq:modham1} and defining $p(r)=1-r^2$ we get an expression for $h'(\alpha_{0,\Omega,d})$ as a differential operator operating on two components $\phi_1$ and $\phi_2$ of a vector in $C$.
\begin{equation}\label{modHamSphere}
h'(\alpha_{0,\Omega,d})\begin{bmatrix}\phi_1\\\phi_2\end{bmatrix}=\frac{\pi}{2}\begin{bmatrix}
    \nabla\cdot(p\nabla\phi_2)-(d-1)\phi_2\\p\phi_1
\end{bmatrix}.
\end{equation}
The eigenvalue problem of $h'(\alpha_{0,\Omega,d})$ therefore becomes the following coupled partial differential equation:
\begin{align}\label{coupEig}
    \begin{split}
        i\frac{2\lambda}{\pi}\phi_{1}&= \nabla\cdot(p\nabla\phi_{2})-(d-1)\phi_{2}\\i\frac{2\lambda}{\pi}\phi_{2}&=p \phi_{1}.
    \end{split}
\end{align}
We uncouple this equation by transforming it into a second order PDE in $\phi_{2}$:
\begin{equation}
    -\frac{4\lambda^2}{\pi^2}\phi_{2}=p^2\nabla^2\phi_{2}+pp'\frac{\partial \phi_{2}}{\partial r}-(d-1)p\phi_{2}.\end{equation}
Since this differential operator is spherically symmetric, we can symmetrize it by considering fields of the form $\phi(\vectorbold{x})=\psi(r)Y_{\ell}^{m,\eta}$ as described in Section \ref{spher}. We get an ODE on the interval $[0,1)$ which depends only on $\ell$, 
\begin{equation}
   \frac{4\lambda^2}{\pi^2}\psi(r)=-(1-r^2)^2 \left(\psi''(r)+\left(\frac{d-1}{r}-\frac{2r}{1-r^2} \right)\psi'(r)-\left(\frac{\ell(\ell+d-2)}{r^2}+\frac{d-1}{1-r^2} \right)\psi(r) \right).
\end{equation}
On the surface, this looks like a Sturm–Liouville equation. But, it is not quite. If we put this into Sturm–Liouville form, the weight function diverges at the boundary, which nullifies an assumption necessary for classical Sturm-Louville theory to apply. There are also no boundary conditions obvious at the boundary from the setup. The only sensible restriction is that the function is smooth inside the sphere and, in particular at $r=0$. This means that the spectrum will be likely continuous. This does match with Casini and Huerta's result that this operator is a Laplacian in some hyperbolic space \cite{Casini_2010}. To take advantage of some existing results in the literature, we make the substitutions $r=\tanh(x)$ and $\psi = \sinh(x)^{\ell}\cosh(x)^{d+\ell-1}\Phi$. We define $\gamma = \ell + \frac{d-1}{2}$,\footnote{For $d=2$, replace the $\ell$ with $|\ell|$ here and elsewhere in the section.}  and rewrite the ODE as
\begin{equation}\label{jacODE}
\Phi''(x)+2\gamma \big(\coth(x)+\tanh(x)  \big) \Phi'(x)+\left(4\gamma^2+(2\frac{\lambda}{\pi})^2 \right)\Phi(x)=0.
\end{equation}
There is also the matter of boundary conditions. Using the Frobenius method, the two independent solutions for $\psi$ behave like $\psi(r)\sim r^\ell$ and $\psi(r)\sim r^{-\ell-d+2}$ near $r=0$. The exception is if $d=2$ and $\ell=0$, in which case the second term is logarithmic. In either case, the smooth solution at the origin of the $n$-ball is the $r^\ell$ solution.\footnote{For confirmation that the resulting functions are analytic at zero, recall that $r^\ell Y_\ell^0(\theta_1,\dots,\theta_{d-1})$ is a polynomial.} This gives a boundary condition of $\Phi(0)=A$ for some $A\neq 0$. If we try to find another boundary condition at $r=1$ by using the Frobenius method again, we see that $\psi(r)\sim (1-r)^{\pm I\lambda }$. So, the solution is oscillatory as it approaches the surface of the sphere. This is what ultimately implies a continuous spectrum as it means $\Phi(x)$ asymptotes to a sinusoidal function as $x\to \infty$. This means we reasonably expect there to be some Fourier-type transform giving some orthogonality relations which would allow us to use the eigenvectors of Equation \eqref{jacODE} as a continuous basis. Fortunately, this continuity of the spectrum, as opposed to the discreteness of a classical 
Sturm–Liouville spectrum, turns out out to be very helpful as it turns what would otherwise be difficult to handle infinite sums into integrals.
 
 Koornwinder \cite{Koornwinder_Schempp_Koornwinder_1984} studied solutions to Equation \eqref{jacODE} with initial condition $\Phi(0)=1$. The solutions are the unnormalized Jacobi functions, which we will label as $\Phi_\gamma(x,\lambda)$.
 The Jacobi functions can be represented in terms of the hypergeometric function as 
 \begin{equation}
     \Phi_\gamma(x,\lambda)=\cosh(x)^{-2\gamma-2i\lambda/\pi }{}_2F_1\left( \gamma+i\frac{\lambda}{\pi},\frac{1}{2}+i\frac{\lambda}{\pi};\gamma+\frac{1}{2};\tanh^2(x) \right).
 \end{equation}

 From Ref.~\cite{Koornwinder_Schempp_Koornwinder_1984}, the Jacobi functions do indeed form the kernel of a Fourier type transform called the Fourier-Jacobi transform. Equations 1.2 and 1.3 from \cite{Koornwinder_Schempp_Koornwinder_1984} imply the following orthogonality relation:
 \begin{equation}\label{oortho}
    \delta(\lambda-\lambda')= \frac{4\lambda\sinh(\lambda)|\Gamma(\gamma+i \frac{\lambda}{\pi})|^2}{\pi^3\Gamma(\gamma+\frac{1}{2})^2}\int_0^\infty \diff x \text{ } \Phi_\gamma(x,\lambda)\Phi_\gamma(x,\lambda')(\sinh(x)\cosh(x))^{2\gamma}.
 \end{equation}

This allows us to write, up to a normalization constant, the second component of the symmetrized eigenvector of $h'(\alpha_{0,\Omega,d})$ after transforming back into radial coordinates. We will label this function $ \psi_{\ell,d}(r,\lambda)$ which can be written as
\begin{equation}\label{psi}
    \psi_{\ell,d}(r,\lambda) = A r^\ell(1-r^2)^{i\lambda/\pi}{}_2F_1 \left(\gamma+i\frac{\lambda}{\pi},\frac{1}{2}+i\frac{\lambda}{\pi};\gamma+\frac{1}{2};r^2 \right).
\end{equation}
It is worth noting that despite the apparent dependence on $i$, this is a real function, due to the Pfaff transformation of the hypergeometric function. These functions satisfy an orthogonality relation following from Equation \eqref{oortho}:
\begin{equation}\label{Ortho1}
    \delta(\lambda-\lambda')= \frac{4\lambda\sinh(\lambda)|\Gamma(\gamma+i \frac{\lambda}{\pi})|^2}{A^2\pi^3\Gamma(\gamma+\frac{1}{2})^2}\int_0^1 \diff r \text{ } r^{d-1}   \frac{\psi_{\ell,d}(r,\lambda) \psi_{\ell,d}(r,\lambda')}{1-r^2}.
\end{equation}
We can easily find the first component of the $\psi_{\ell,d,1}$ eigenvector using the second relation in Equation \eqref{coupEig} allowing us to write the write the eigenvectors of $f(\alpha)$ as
\begin{equation}\label{eigs}
    \phi_{d,\ell,m,\lambda}(\vectorbold{x}) = \begin{bmatrix}
        \phi^1_{d,\ell,m,\lambda}(\vectorbold{x})\\ \phi^2_{d,\ell,m,\lambda}(\vectorbold{x})
    \end{bmatrix}
    = \psi_{\ell,d}(r,\lambda)Y_{\ell}^{m,\eta}(\theta_1,\dots,\theta_{d-1})\begin{bmatrix}
        \frac{2\lambda i}{\pi(1-r^2)} \\ 1
    \end{bmatrix}.
\end{equation}
To normalize the eigenvectors, we insert them into a continuous version of the orthogonality relation in Equation \eqref{Normalization} in Appendix \ref{sec:modHamLatt},
\begin{align}\label{Ortho2}
    \begin{split}
    \delta_{\ell',\ell}\delta_{m',m}\delta_{\eta',\eta}\delta(\lambda-\lambda')&=-i\lambda' \int_\Omega  \diff^d \vectorbold{x} \text{ }\phi^2_{d,\ell',m',\lambda'}(\vectorbold{x})^*\phi^1_{d,\ell,m,\lambda}(\vectorbold{x})-\phi^1_{d,\ell',m',\lambda}(\vectorbold{x})^*\phi^2_{d,\ell,m,\lambda}(\vectorbold{x})\\
    \implies \delta_{\ell',\ell}\delta_{m',m} \delta_{\eta',\eta}\delta(\lambda-\lambda')&= \frac{4\lambda^2}{\pi}\int_{\partial \Omega} \diff^{d-1} \vectorbold{S} \text{ } Y_{\ell'}^{m',\eta'*}Y_{\ell}^{m,\eta}\int_0^1 \diff r  \text{ } r^{d-1} \frac{\psi_{\ell,d}(r,\lambda')\psi_{\ell,d}(r,\lambda)}{1-r^2}.
    \end{split}
\end{align}

Now, using the orthogonality relation for spherical harmonics and comparing Equations \eqref{Ortho1} and \eqref{Ortho2} we can solve for the normalization constant $A$ as
\begin{equation}\label{ACon}
    A = \frac{\sqrt{\sinh(\lambda)}|\Gamma(\gamma+i \frac{\lambda}{\pi})|}{\sqrt{\lambda} \pi\Gamma(\gamma+\frac{1}{2})}.
\end{equation}
We have now completely described the normalized eigenvectors of $h'(\alpha_{0,\Omega,d})$ through Equations \eqref{psi}, \eqref{eigs}, and \eqref{ACon}. 

\subsection{The Eigenbasis of the Skewed Vacuum Correlation Matrix}
Now that we have the spectral decomposition of $\alpha_{0,\Omega,d}$, we can extend Equation \eqref{eq:alphaBasis} to transform a matrix into the $\alpha_{0,\Omega,d}$ eigenbasis.

That is, given an arbitrary matrix $A$ in $C^2$ with $2\times2$ matrix valued kernel $K(\vectorbold{x},\vectorbold{y})$, we define \textit{the kernel of $A$ with respect to $\alpha_{0,\Omega,d}$} or alternatively \textit{the kernel of $A$ in the $\alpha_{0,\Omega,d}$ eigenbasis} as
\begin{equation}\label{Kernelalpha}
    \tilde{K}_{\ell,m,\eta,\ell',m',\eta'}(\lambda,\lambda') = -i \lambda'\int\diff\vectorbold{x}\int\diff\vectorbold{y}\text{ } \psi_{\ell,d}(r,\lambda)\psi_{\ell',d}(s,\lambda')  Y_{\ell}^{m,\eta}Y_{\ell'}^{m',\eta'}\begin{bmatrix}
         1&\frac{2\lambda' i}{\pi(1-s^2)}\end{bmatrix}K(\vectorbold{x},\vectorbold{y})\begin{bmatrix}
        \frac{2\lambda i}{\pi(1-r^2)} \\ 1
    \end{bmatrix}.
\end{equation}
Note that placing a matrix in the $\alpha_{0,\Omega,d}$ eigenbasis also gives the decomposition of the matrix into the spherical wave mode blocks outline in Appendix \ref{spher}.
Exactly as the name suggests, this expression is equivalent to our original kernel up to an isomorphism of vector spaces. In particular, we can extend Equation \ref{eq:DiagonalResolvent} to compute the kernel  $\tilde{R}_{z,\ell,m,\eta,\ell',m',\eta'}(\lambda,\lambda')$ of the resolvent of $\alpha_{0,\Omega,d}$ in the $\alpha_{0,\Omega,d}$ eigenbasis.
\begin{equation}\label{eq:DiagonalResolventQFT}
    \tilde{R}_{z,\ell,m,\eta,\ell',m',\eta'}(\lambda,\lambda')= \frac{\delta(\lambda-\lambda')\delta_{\eta,\eta'}\delta_{\ell,\ell'}\delta_{m,m'} }{z+i\coth(\lambda_\mu)}.
    \end{equation}
Now that we have used conformal invariance to completely understand the spectrum of the $\alpha_{0,\Omega,d}$, we can perturb it to construct series for the entropy and mutual information of nearby states. We describe this in the next section.

\section{Perturbing a Scalar Massless Vacuum on a Sphere }\label{petuebSphere}

In this section, we give a general procedure for expanding around the vacuum of a massless scalar on a sphere, which will be applicable to both cases we consider in this paper: the mutual information of distant spheres and the entropy difference of low temperatures states. 

Given the open unit sphere $\Omega$, define $C_\Omega$ as its configuration space: the set of tempered distributions on the unit sphere. Let $\alpha(t)$ be a one parameter family of skewed correlation matrices on $C_{\Omega}^2$ with a power series representation $\alpha(t)=\sum_{k=0}^\infty \alpha^{(k)} t^k$ that satisfies $\alpha(0)=\alpha_{0,\Omega,d}$, the skewed correlation matrix of the $d$-dimensional vacuum on a sphere. Here, the $\alpha^{(k)}$ are a sequence\footnote{We label this sequence with a superscript $(k)$, where brackets are used to distinguish from powers. }  of matrices on the sphere such that $\alpha^{(0)}=\alpha_{0,\Omega,d}$. In particular, this allows us to write the difference $\Delta\alpha = \alpha(t)-\alpha(0)$ as a power series $\Delta\alpha=\sum_{k=1}^\infty \alpha^{(k)} t^k$.

Before we begin, we will check that the assumptions outlined in Section \ref{smooth} which allow us to take the trace over $L^2(\Omega)^2$. Clearly, $\Omega$ is bounded. We will choose to make the assumption that the entries of the kernel of $\Delta \alpha$ are smooth, as that will be true for both cases we look at in this paper. Lastly, we note that $\alpha_{0,\Omega,d}$ is densely defined on $L^2(\Omega)^2$. For example, we can set the domain of definitions of $\alpha_{0,\Omega,d}$ to be set of smooth functions on $\Omega$ with compact support. To confirm this, note that $X_{0,\Omega,d}$ is already a bounded operator on $L^2(\Omega)^2$ \cite{compactSource}. Furthermore, since a function on the domain of definition will be zero on some neighborhood of the boundary, it avoids the discontinuity discussed in Section \ref{fieldBlocks} and its image under $P_{0,\Omega,d}$ is bounded and therefore in $L^2(\Omega)^2$ meaning $\alpha_{0,\Omega,d}$ is indeed densely defined.

For cleanliness, in this section we will label the wave mode subspaces discussed in Appendix \ref{spher} with a single index $\upsilon=(\ell,m,\eta)$. Suppose that we know the form of the block matrix of $\alpha^{(k)}$ from $\upsilon_1$ to $\upsilon_2$, which we label as $\alpha^{(k)}_{\upsilon_1,\upsilon_2}$. In particular, suppose we know its form in the $\alpha_{0,\Omega,d}$ basis, as defined in Equation \eqref{Kernelalpha}. We label the symmetrized components of the $\alpha^{(k)}$ in this basis by $\tilde{K}^{(k)}_{\upsilon_1,\upsilon_2}(\lambda,\lambda')$.

Given the matrix function $h(t)=\frac12 \arccot(t)$, our goal is to compute the power series representation for the entropy $S(t)=\Trace_{L^2(\Omega)^2}[h(\alpha(t))]=\sum_{N=0}^\infty P_N t^N$ in $t$. We begin with the quantum field theoretic version of Equation \eqref{eq:perturbedNice},
\begin{equation}
   S(t)= S(0) + \Trace_{L^2(\Omega)^2}\left[h'(\alpha_{0,\Omega,d})\Delta\alpha\right]+\sum_{n=2}^\infty\frac{1}{2\pi n i }\oint_\xi \diff z \text{ }h'(z)\Trace_{L^2(\Omega)^2}\left[\left[(z-\alpha_{0,\Omega,d})^{-1}\Delta\alpha\right]^{n}\right].
\end{equation}
We now substitute in our power series representation for $\Delta\alpha(t)$,
\begin{align}
    \begin{split}
        S(t) &=S(0)+ \Trace_{L^2(\Omega)^2}\left[h'(\alpha_{0,\Omega,d})\Delta\alpha\right]\\&+\sum_{n=2}^\infty\frac{1}{2\pi n i }\oint_\xi \diff z \text{ }h'(z)\Trace_{L^2(\Omega)^2}\left[\left[(z-\alpha_{0,\Omega,d})^{-1}\left(\sum_{k=1}^\infty \alpha^{(k)} t^k\right)\right]^{n}\right].
    \end{split}
\end{align}
    
Suppose that $\Trace_{C_{\Omega}^2}\left[h'(\alpha_{0,\Omega,d})\Delta\alpha\right]$ has a power series representation $\sum_{N=1}^\infty Q_Nt^N$. We can see that the term of order $t^N$ will have a single contribution of $Q_N$ from $\Trace_{C_{\Omega}^2}\left[h'(\alpha_{0,\Omega,d})\Delta\alpha\right]$ and the remaining contributions come from all possible ways to sum a set of positive integers $k_1,\dots,k_n$ to $N$ where $n\geq 2$:
\begin{equation}
    P_N -Q_N= \sum_{\substack{N=\sum_{j=1}^nk_j \\n\geq 2,\text{ } k_j\geq 1}}\frac{1}{2\pi n i }\oint_\xi \diff z \text{ }h'(z)\Trace_{C_{\Omega}^2}\left[\prod_{i=1}^n \left(z-\alpha_{0,\Omega,d})^{-1}(\alpha^{(k_j)}\right)\right].
\end{equation}
Furthermore, each of these terms can be subdivided into more terms by splitting $\alpha_{0,\Omega,d}$ and the $\alpha^{(k)}$ into wave mode block matrices. As described in Section \ref{Blockpath}, the block matrices of their products can be given by a sum of paths over the wave mode subspaces $C_\upsilon^2$. Since we are taking a trace, we only need to include the paths that begin and end at the same subspace. Each term contributing to $P_N$ therefore corresponds to a sequence of tuples $(k,\upsilon)$ with length greater than one, written schematically as
\begin{equation}
    (k_1,\upsilon_1)\to (k_2,\upsilon_2)\to\dots\to(k_n,\upsilon_n).
\end{equation}
This sequences generates a sequence of symmetrized matrices, namely
\begin{equation}
    \alpha^{(k_1)}_{\upsilon_1,\upsilon_2},\alpha^{(k_2)}_{\upsilon_2,\upsilon_3},\dots,\alpha^{(k_n)}_{\upsilon_n,\upsilon_1}.
\end{equation}
Practicing a bit of foresight, it will be very helpful to group the terms which will produce the same contribution due the invariance of the trace under cyclic permutations. To avoid listing the same term twice, we need the following definition. A \textit{circular list} is an ordered list under the equivalence relation of cyclic equivalence. We take circular lists to be periodic with $x_{j}=x_{n+j}$ for some length $n$. 

Given a circular list $\tau$, we define $S(\tau)$ to be the number of cyclic permutations that leave $\tau$ unchanged. Equivalently, $S(\tau)$ is the number of equal ordered lists $\tau$ can be sliced into.\footnote{To demonstrate the idea, here are some examples of $S$ for circular lists of integers: $S(1,1,1,1)=4$, $S(1,0,1,0)=2$, and $S(1,1,1,0)=1$.} We denote by $\chi_N$ the set of all circular lists of tuples $ (k_i,\upsilon_i)$ such that $\sum_i k_i =N$.  We can now write the $t^N$ term using a sum over $\chi_N$ as
\begin{equation}
    P_N -Q_N= \sum_{\tau\in\chi_N}\frac{1}{2S(\tau)\pi  i }\oint_\xi \diff z \text{ }h'(z)\Trace_{C_{\Omega}^2}\left[\prod_{i=1}^n \left((z-\alpha_{0,\Omega,d})^{-1}\alpha_{\upsilon_j,\upsilon_{j+1}}^{(
    k_j)}\right)\right].
\end{equation}
The $\frac{1}{S(\tau)}$ prevents the sum from over-counting the cyclically equivalent terms. We now find the kernel of $\prod_{i=1}^n \left(\alpha_{\upsilon_j,\upsilon_{j+1}}^{k_j}(z-\alpha_{0,\Omega,d})^{-1}\right)$ in the $\alpha_{0,\Omega,d}$ basis to be
\begin{equation}
    K_{\tau}(\lambda_1,\lambda_{n+1})=\left[\prod_{j=2}^{n} \int_{-\infty}^{\infty}\diff \lambda_{j}\right]\frac{\prod_{j=1}^{n} \tilde{K}_{\upsilon_{j},\upsilon_{j+1}}^{(k_{j})}(\lambda_{j},\lambda_{j+1})}{\prod_{j=1}^{n} (z+i\coth(\lambda_{j}))}. 
\end{equation}
To take the trace, we set $\lambda_{n+1}=\lambda_1$ and integrate over $\lambda_1$. Therefore, the coefficients of the expansion can be written in terms of iterated integrals as
\begin{equation}
   P_N -Q_N =\sum_{\tau \in \chi_N } \frac{1}{2S(\tau,\chi)\pi i}\left[\prod_{j=1}^{n} \int_{-\infty}^{\infty}\diff \lambda_{j}\right]\oint_\xi \diff z \text{ }\frac{h'(z)\tilde{K}_{\upsilon_{n},\upsilon_{1}}^{(k_{n})}(\lambda_n,\lambda_{1})\prod_{j=1}^{n-1}\tilde{K}_{\upsilon_{j},\upsilon_{j+1}}^{(k_{j})}(\lambda_{j},\lambda_{j+1})}{\prod_{j=1}^{n} (z+i\coth(\lambda_{j}))}.
\end{equation}
Recall that the contour integral encloses the portion of the imaginary axis such that $|z|>1$. Now, we can use the Cauchy integral theorem to evaluate the contour integral.
Our final form is now 
\begin{align}\label{genExpand}
    \begin{split}
    P_N-Q_N=\sum_{\tau \in \chi_N}\frac{1}{S(\tau,\chi)}\left[\prod_{j=1}^{n} \int_{-\infty}^{\infty}\diff \lambda_{j}\right]&\tilde{K}_{\upsilon_{n},\upsilon_1}^{(k_n)}(\lambda_n,\lambda_{1})\prod_{j=1}^{n-1}\tilde{K}_{\upsilon_{j},\upsilon_{j+1}}^{(k_{j})}(\lambda_{j},\lambda_{j+1})\\&\times\sum_{q=1}^{n}\left[ \frac{i^{n-1}h'(-i\coth(\lambda_{q}))}{\prod_{j=1,j\neq q}^{n} (\coth(\lambda_{q})-\coth(\lambda_{k}))}\right].
    \end{split}
\end{align}
We note that in general, there may be infinite number of terms to compute each $P_N$. In the cases we consider later, the sum reduces to a finite number of nonzero terms.
We therefore write the entropy as 
\begin{equation}
    S(t)=S(0)+\sum_{N=1}^\infty P_N t^N
\end{equation}

\section{Mutual Information of Identical Spheres Separated by $R$}
\label{MISect}

Consider two spheres $\Omega_1$ and $\Omega_2$ of unit radius whose centres are separated by a distance $R$. In particular, our goal is to find an expansion for the mutual information between these two spheres in terms of their inverse separation $\frac{1}{R}$ for large $R$. After making a simplification owing to the fact that the spheres are identical, we will expand the skewed correlation matrix in terms of $\frac{1}{R}$. We can then use the general procedure of Section \ref{petuebSphere} to compute the full expansion and lowest order term of the mutual information.

If $R>2$, the spheres do not intersect and we can decompose the configuration space as a direct sum $C=C_{1}\oplus C_{2}$, where $C$ is the configuration space on $\Omega_1\cup \Omega_2$, $C_1$ is the configuration space on $\Omega_1$ and $C_2$ is the configuration space on $\Omega_2$. Note that $C_1$ and $C_2$ are both isomorphic to $C_{\Omega}$, the configuration space for a scalar field on the unit sphere. We use the result of Section \ref{sec:Ent/MI} to find the mutual information. If we set up spherical coordinates on the spheres such that the central axes of the spheres point towards each other, the problem has reflection symmetry. Therefore, in the proper basis, we have a simplified form for the correlation matrix $\alpha$,
\begin{equation}
    \alpha = \begin{bmatrix}\alpha_0&\alpha_1\\\alpha_1 &\alpha_0.
    \end{bmatrix}
\end{equation}
Here $\alpha_0$ is the correlation matrix of a sphere of radius $1$. With correct choices of coordinates systems, this is precisely $\alpha_{0,\Omega,d}$, the operator analyzed in Section \ref{sec:Diagsphere}.
We can now write the mutual information as
\begin{equation}
    I = \Trace_{C^2}\left[h\left(\begin{bmatrix}\alpha_0&0\\0 &\alpha_0
    \end{bmatrix}\right)-h\left(\begin{bmatrix}\alpha_0&\alpha_1\\\alpha_1 &\alpha_0
    \end{bmatrix}\right)\right].
\end{equation}
We can compute the mutual information using Equation \eqref{perturbedNice} , using $h(t)=\frac{t}{2}\arccot(t)+\frac{1}{4}\ln(-\frac{1}{4}(t^2+1))$, as
\begin{align}
\begin{split}
    I =& -\Trace_{L^2(\Omega)^2}\left[h'\left(\begin{bmatrix}\alpha_0&0\\0 &\alpha_0
    \end{bmatrix}\right)\begin{bmatrix}0&\alpha_1\\\alpha_1&0
    \end{bmatrix}\right]\\&-\sum_{n=1}^\infty\frac{1}{2\pi n i}\oint_\xi \diff z \text{ } h'(z)\Trace_{L^2(\Omega)^2}\left[ \left[\begin{bmatrix}0&\alpha_1\\\alpha_1&0
    \end{bmatrix}(z-\begin{bmatrix}\alpha_0&0\\0 &\alpha_0
    \end{bmatrix})^{-1}\right]^n\right].
    \end{split}
\end{align}
After multiplying out the block matrices and simplifying, we find that the first term contributes nothing. For the sum, we find the odd $n$ terms in the sum don't contribute and the even $n$ terms can be simplified into essentially the same form as Equation \eqref{perturbedNice},
\begin{equation}\label{perturbedMI}
    I = -\frac{2}{2\pi i}\sum_{n=1}^\infty\oint_\xi \diff z \text{ } h'(z)\Trace_{L^2(\Omega)^2}\left[ \left[\alpha_1(z-\alpha_0)^{-1}\right]^{2n}\right].
\end{equation}
This allows us to apply the results of Section \ref{petuebSphere} directly to $\alpha_1$ instead of to $ \begin{bmatrix}0&\alpha_1\\\alpha_1 &0
    \end{bmatrix}$. 
    
    Note that it would still be possible to use our approach to analyze two spheres of different sizes. But this final simplification would not be possible and we would need to work with the full forms of the skewed correlation matrices the entire way through. Regardless, the case of spheres with different radii is related to the case where the spheres have equal radii by conformal symmetry.

\subsection{Expanding and Symmetrizing $\alpha_1$}

Now we need to find an expansion for $\alpha_1$ in terms of $\frac{1}{R}$. For concreteness,  let us place one sphere, called $\Omega_1$, centered on the origin with radius $1$ and place another, called $\Omega_2$, centered on the point $\vectorbold{R}=(R,0,\dots,0)$ with radius $1$. Then we write $\alpha_1$ as a linear map from functions on $\Omega_1$ to functions on $\Omega_2$. This operator has the following $2\times 2$ matrix valued kernel, which can be found by restricting the domain and image of Equation \eqref{vacCorrs},
\begin{equation}
    K_1(\vectorbold{x},\vectorbold{y}) =  -\frac{\Gamma(\frac{d-1}{2})}{2\pi^{\frac{d+1}{2}}}\begin{bmatrix}
      0&\frac{d-1}{|\vectorbold{x}-\vectorbold{y}|^{d+1}}\\ \frac{1}{|\vectorbold{x}-\vectorbold{y}|^{d-1}}&0
    \end{bmatrix}.
\end{equation}
Here $\vectorbold{x}\in \Omega_1$ and $\vectorbold{y}\in \Omega_2$. We note that since $\Omega_1$ and $\Omega_2$ don't share a boundary, this kernel has smooth entries and so we can comfortably use the expansion in Section \ref{petuebSphere}. We re-parameterize $\vectorbold{y}$ so that it is centered on its sphere and its positive $x$ direction is pointing towards the sphere at the origin, $\vectorbold{y}\to\vectorbold{R}-\vectorbold{y}$ giving us
\begin{equation}
    K_1(\vectorbold{x},\vectorbold{y}) =  -\frac{\Gamma(\frac{d-1}{2})}{2\pi^{\frac{d+1}{2}}}\begin{bmatrix}
      0&\frac{d-1}{|\vectorbold{R}-(\vectorbold{y}+\vectorbold{x})|^{d+1}}\\ \frac{1}{|\vectorbold{R}-(\vectorbold{y}+\vectorbold{x})|^{d-1}}&0
    \end{bmatrix}.
\end{equation}
If we define the function $H^{a}_k(\vectorbold{x})=r^kC^{a}_k(\cos(\theta_1))$, where $C^{a}_k$ denotes the Geigenbaur polynomials described in Section \ref{highSim} using spherical coordinates, we can write $K$ in terms of an infinite series in $\frac{1}{R}$,
\begin{equation}\label{kdel}
    K_1(\vectorbold{x},\vectorbold{y}) =-\frac{\Gamma(\frac{d-1}{2})}{2\pi^{\frac{d+1}{2}}R^{d-1}}\sum_{k=0}^\infty\frac{1}{R^k}
    \begin{bmatrix}
      0&(d-1)u(k-2)H^{\frac{d+1}{2}}_{k-2}(\vectorbold{y}+\vectorbold{x})\\ H^{\frac{d-1}{2}}_k(\vectorbold{y}+\vectorbold{x})&0
    \end{bmatrix}.
\end{equation}
Here $u(n)$ is the Heaviside Step function, which is equal to $0$ for negative integers and $1$ for non-negative integers.

It will be convenient to decompose each term in the expansion of $\alpha_1$ into spherical wave modes. Notice that $\alpha_1$ is symmetric with respect to the representation of $SO(d-1)$ corresponding to rotations around $\vectorbold{R}$. As discussed in Section \ref{spher}, this imposes selection rules on the spherical block matrices. We will label the symmetrized block matrices of the $k$-th term of Equation \eqref{kdel} in $d$ spatial dimensions by $\alpha_{\ell_1,\ell_2,m}^{(k),d}$ and its $2\times 2$ radial kernel by $K_{\ell_1,\ell_2,m}^{k,d}(r,s)$. To compute $K_{\ell_1,\ell_2,m}^{k,d}$, we need to evaluate the  following integral for $0<r,s<1$,
\begin{equation}\label{intH}
    \int_{|\vectorbold{x}|=1} \diff^d \vectorbold{x}\int_{|\vectorbold{y}|=1} \diff^d  \vectorbold{y} \text{ }H^{\alpha}_k(s\vectorbold{y}+r\vectorbold{x})Y_{\ell_1}^{m,0}(\vectorbold{x})Y_{\ell_2}^{m,0}(\vectorbold{y})^*.
\end{equation}
We will not solve this integral in general. However, we will describe an algorithm to compute an equivalent expansion for $H^{a}_k$ directly.
We will show that this general expansion is equivalent to a finite dimensional linear algebra problem. We begin by rewriting $ H^{a}_k(\vectorbold{x}+\vectorbold{y})$ in terms of solid harmonics using the connection formula for Gegenbauer polynomials given by Equation 18.18.16 in NIST's Digital Library of Mathematical Functions \cite{NIST:DLMF},
\begin{align}\label{Hdecomp1}
    \begin{split}
       H^{a}_k(\vectorbold{x})&=r^kC^{a}_k(\cos(\theta_1))\\&=r^k\sum_{j=0}^{\lfloor\frac{k}{2}\rfloor} D_{d,k,j,a}C^{\frac{d}{2}-1}_{k-2j}(\cos(\theta_1))
       \\&=\sum_{j=0}^{\lfloor\frac{k}{2}\rfloor}|\vectorbold{x}|^{2j} \frac{D_{d,k,j,a}}{B_{d,k-2j}}R_{k-2j}^{0,0}(\vectorbold{x}).
    \end{split}
\end{align}
Comparing with Equation \eqref{homBasis}, we see that $H^{\alpha}_k$ is a homogeneous polynomial of degree $k$.\footnote{Note that this derivation only works as written  for $d\geq 3$.  For $d=2$, we would also need to make use of Equation \eqref{lim2D} as well as the fact that $\lim_{d\to 2}\frac{D_{d,k,j,a}}{B_{d,k-2j}}$ is a nonzero constant. But we do still conclude that $H^{a}_k$ is a homogeneous polynomial of degree $k$.} Here we have defined a constant $D_{d,k,j,\alpha}$ and use the constant $B_{d,\alpha}$ defined in Equation \eqref{B},
\begin{equation}\label{Dconst}
    D_{d,k,j,a} = \frac{(\frac{d}{2}-1+k-2j)\Gamma(a+k-j)\Gamma(\frac{d}{2})\Gamma(a -\frac{d}{2}+1+\ell)}{j!(\frac{d}{2}-1)\Gamma(a)\Gamma(\frac{d}{2}+k-j)\Gamma(a -\frac{d}{2}+1)}.
\end{equation}
Since $H^{a}_k(\vectorbold{x})$ is a homogeneous polynomial of degree $k$, $H^{a}_k(\vectorbold{x}+\vectorbold{y})$ will be a homogeneous polynomial in both $\vectorbold{x}$ and $\vectorbold{y}$ such that the total degree of in $\vectorbold{x}$ and $\vectorbold{y}$ combined is $k$. Therefore, we can expand in terms of the basis functions $G_{\ell,j}^{m,\eta}$ defined in Equation \eqref{homBasis},\footnote{As written, this sum only works for $d\geq 3$. For $d=2$, $\ell$ can be negative and that must be accounted for in the indices of the sum. Once again, this does not change the important outcome, the fact that the symmetrization can be computed with a linear system of equations.}
\begin{align}
\begin{split}
H^{a}_k(\vectorbold{x}+\vectorbold{y})&= \sum_{n,m,k_1,k_2,j_1,j_2,\eta}A^{k,a,d}_{n,m,j_1,j_2} G_{n,j_1}^{m,\eta}(\vectorbold{x})^*G_{k-n,j_2}^{m,\eta}(\vectorbold{y})\\
&= \sum_{n,m,k_1,k_2,j_1,j_2,\eta}A^{k,a,d}_{n,m,j_1,j_2} |\vectorbold{x}|^{n_1}|\vectorbold{y}|^{k-n} Y_{n-2j_1}^{m,\eta}\left(\frac{\vectorbold x}{|\vectorbold x|}\right)^*Y_{k-n-2j_2}^{m,\eta}\left(\frac{\vectorbold y}{|\vectorbold y|}\right).
\end{split}
\end{align}
The first line of this equation writes the homogeneous polynomial on the left hand side as a linear combination of all possible combinations of homogeneous polynomials $G$ with the same degree on the right hand side. But the homogeneous polynomials are a finite dimensional vector space, meaning that the constants $A^{k,a,d}_{n,m,j_1,j_2}$ can be computed through a straightforward finite dimensional linear system of equations\footnote{In the way described here, this is a linear algebra problem over the field of quadratic irrationals. This is only for cleanliness. By omitting normalization constants, we can turn it into a problem over the (complex) rationals.}. Here we note that for these coefficients, the indices $k$ and $a$ are indices of the Gegenbauer polynomials, the index $d$ corresponds to the spatial dimension, while $n,m,j_1,j_2$ label a basis for homogeneous polynomials. There is likely no neat formula for these coefficients in general. However, we can work with special cases. For example, we easily compute the expansion for $k=0$ later and there are likely more universal terms that appear in the expansion for every dimension. But practically, if one wants to compute this decomposition for any specific dimension, say $d=3$, they can quickly find the $A^{k,a,d}_{n,m,j_1,j_2}$ exactly by solving this linear algebra problem. The only requirement is knowing the Cartesian form of the spherical harmonics for that dimension so that the $G$ polynomials can be written.

This expansion gives some additional selection rules for $K_{\ell_1,\ell_2,m}^k$. By noting that the degrees of the spherical harmonics are of the form $\ell_1=n_1-2j_1$ and $\ell_2=k-n_1-2j_2$, we see that $\ell_1+\ell_2\leq k$ and $\ell_1-\ell_2 \equiv k \mod 2$.\footnote{For $d=2$, the first selection rule is $|\ell_1|+|\ell_2|\leq k$.} 
Now, let us define two new sets of constants
by re-indexing $A^{k,\alpha,d}_{n,m,j_1,j_2}$ and factoring in the other constants in Equation \eqref{kdel},
\begin{align}
\begin{split}
    F^{X,k,j,d}_{\ell_1,\ell_2,m}&= \frac{\Gamma(\frac{d-1}{2})}{2\pi^{\frac{d+1}{2}}} A^{k,\frac{d-1}2,d}_{\ell_1+2j,m,j,\frac{k-\ell_1-\ell_2}{2}-j}\\
    F^{X,k,j,d}_{\ell_1,\ell_2,m}&= \frac{\Gamma(\frac{d+1}{2})}{\pi^{\frac{d+1}{2}}} A^{k-2,\frac{d+1}2,d}_{\ell_1+2j,m,j,\frac{k-\ell_1-\ell_2}{2}-j-1}
\end{split}
\end{align}
Now, by grouping terms with matching spherical harmonics, we can now write the general form of $K_{\ell_1,\ell_2,m}^{k,d}$ as
\begin{equation}\label{Kgenform}
K_{\ell_1,\ell_2,m}^{k,d}(s,r)=-\begin{bmatrix}0&u(k-2)\sum_{j=0}^{\frac{k-\ell_1-\ell_2-2}{2}}F^{P,k,j,d}_{\ell_1,\ell_2,m}s^{\ell_1+2j}r^{k-\ell_1-2j-2}\\\sum_{j=0}^{\frac{k-\ell_1-\ell_2}{2}}F^{X,k,j,d}_{\ell_1,\ell_2,m}s^{\ell_1+2j}r^{k-\ell_1-2j} &0
    \end{bmatrix}.
\end{equation}

For clarity, we remind the reader that $d$ is the spatial dimension, $\ell_1$ denotes the $SO(d)$ index of the preimage subspace,  $\ell_2$ denotes the $SO(d)$ index of the image wave mode subspace, $m$ denotes their shared $SO(d-1)$ wave mode index guaranteed by Schur's Lemma, and $k$ means that we are looking at the order $\frac{1}{R^k}$ term of $\Delta\alpha$.
\subsection{The Symmetrized Kernel with respect to $\alpha_{0,\Omega,d}$}
We compute the symmetrized kernel in the basis of eigenvectors of $\alpha_{0,\Omega,d}$, labelled as $\tilde{K}_{\ell_1,\ell_2,m}^{k,d}$. We substitute Equation \eqref{Kgenform} into the definition of $\tilde{K}$ given in Equation \eqref{Kernelalpha} using the eigenvectors of $\alpha_{0,\Omega,d}$ described in Equation \eqref{eigs},
\begin{align}\label{Koper}
\begin{split}
   \tilde{K}_{\ell_1,\ell_2,m}^{k,d}(\lambda,\lambda')
&=-\frac{4i\lambda'^2\lambda}{\pi^2}\sum_{j=0}^{\frac{k-\ell_1-\ell_2}{2}}F^{X,k,j}_{\ell_1,\ell_2,m}Q^{j}_{d,\ell_1,0}(\lambda)Q^{\frac{k-\ell_1-\ell_2}2-j}_{d,\ell_2,0}(\lambda')\\&+i\lambda'u(k-2)\sum_{j=0}^{\frac{k-\ell_1-\ell_2-2}{2}}F^{P,k,j}_{\ell_1,\ell_2,m}Q^{j}_{d,\ell_1,1}(\lambda)Q^{\frac{k-\ell_1-\ell_2}{2}-j-1}_{d,\ell_2,1}(\lambda').
\end{split}
\end{align}
Here, we define a function $Q^\nu_{d,\ell,\delta}$, where we recall the definition of $\psi_{\ell,d}$ in Equation \eqref{psi} and our notation that $\gamma=\frac{d-1}2 +\ell$,
\begin{align}
\begin{split}
    Q^\nu_{d,\ell,\delta}(\lambda)&=\int_0^1 \diff r \text{ } \frac{r^{d-1+2\nu+\ell}}{(1-r^2)^{1-\delta}}\psi_{\ell,d}(r,\lambda)\\
    &=\frac{\sqrt{\sinh(\lambda)}|\Gamma(\gamma+i \frac{\lambda}{\pi})|}{\sqrt{\lambda} \pi\Gamma(\gamma+\frac{1}{2})}\int_0^1 \diff r \text{ } r^{2\gamma+2\nu}(1-r^2)^{\delta-1+i\lambda/\pi}{}_2F_1\left(\gamma+i\frac{\lambda}{\pi},\frac{1}{2}+i\frac{\lambda}{\pi};\gamma+\frac{1}{2};r^2 \right).
\end{split}
\end{align}
Now, we make the substitution $u=r^2$,
\begin{equation}   Q^\nu_{d,\ell,\delta}(\lambda)=\frac{\sqrt{\sinh(\lambda)}|\Gamma(\gamma+i \frac{\lambda}{\pi})|}{2\sqrt{\lambda} \pi\Gamma(\gamma+\frac{1}{2})}\int_0^1 \diff u \text{ } u^{\gamma+\nu-\frac12}(1-u)^{-1+\delta+i\lambda/\pi}{}_2F_1\left(\gamma+i\frac{\lambda}{\pi},\frac{1}{2}+i\frac{\lambda}{\pi};\gamma+\frac{1}{2};u\right).
\end{equation}
Now using 7.512.5 in Gradshteyn and Ryzhik \cite{2014776}, we can evaluate this integral using a hypergeometric function,
\begin{align}
    \begin{split}
     Q^\nu_{d,\ell,\delta}(\lambda)&=\frac{\sqrt{\sinh(\lambda)}\Gamma(\gamma+\nu+\frac{1}{2})\Gamma(\delta+i\frac{\lambda}{\pi})|\Gamma(\gamma+i \frac{\lambda}{\pi})|}{2\sqrt{\lambda} \pi\Gamma(\gamma+\frac{1}{2})\Gamma(\gamma+\nu+\frac{1}{2}+\delta+i\frac{\lambda}{\pi})}\\&\times{}_3F_2\left(\gamma+i\frac{\lambda}{\pi},\frac{1}{2}+i\frac{\lambda}{\pi},\gamma+\nu+\frac12;\gamma+\frac{1}{2},\gamma+\nu+\frac{1}{2}+\delta+i\frac{\lambda}{\pi};1\right).
    \end{split}
\end{align}
First we note that we can evaluate the ${}_3F_2$ hypergeometric function in terms of ${}_2F_1$ hypergeometric functions. To do so, we note that $\nu$ is always an integer, and so we can use 16.3.1 and 16.3.3 in the Digital Library of Mathematical Functions \cite{NIST:DLMF}. We can then use Gauss's Summation theorem for hypergeometric function ${}_2F_1$ to rewrite this function in terms of the Gamma function as
\begin{align}\label{Qfun}
    \begin{split}
     Q^\nu_{d,\ell,\delta}(\lambda)&=\frac{\sqrt{\sinh(\lambda)}\Gamma(\delta+i\frac{\lambda}{\pi})\Gamma(\gamma+\nu+\frac{1}{2})|\Gamma(\gamma+i \frac{\lambda}{\pi})|}{2\sqrt{\lambda} \pi\Gamma(\gamma+\frac{1}{2})\Gamma(\gamma+i\frac{\lambda}{\pi})\Gamma(\frac{1}{2}+i\frac{\lambda}{\pi})\Gamma(\gamma+\nu+\delta)\Gamma(\delta+\nu+\frac{1}{2})}\\&\times\sum_{n=0}^{\nu}\frac{\Gamma(\gamma+n+i\frac{\lambda}{\pi})\Gamma(\frac{1}{2}+n+i\frac{\lambda}{\pi})\Gamma(\delta+\nu-n-i\frac{\lambda}{\pi})}{\Gamma(\gamma+\frac{1}{2}+n)}{\nu \choose n}.
    \end{split}
\end{align}
 Once the finite dimensional linear algebra problem finding the constants $F^{P,k,j}_{\ell_1,\ell_2,m}$ and $F^{X,k,j}_{\ell_1,\ell_2,m}$ has been solved, one can use Equation \eqref{Qfun} to write $\tilde{K}_{\ell_1,\ell_2,m}^{k,d}$ in terms of algebraic combinations of Gamma functions.

\subsection{The General Expansion for Mutual Information}
We can now describe the general form of the $\frac{1}{R^N}$ term of the expansion for mutual information using the results of Section \ref{petuebSphere}.
 
Recall the definition of a circular list as an ordered list under the equivalence relation of cyclic equivalence.To properly parameterize the terms, we need the following definitions. Given spatial dimension $d$ and exponent in the  $\frac{1}{R}$ expansion $N$, define $\chi_{d,N}$ to be the set of all circular lists of tuples of the form $(k,\ell,m)$ such that its length $n$ is even, $\sum_{j=1}^n (d-1 + k_j) = N$,  
 $|\ell_j|+|\ell_{j+1}|\leq k_j$, $\ell_j-\ell_{j+1}\equiv k_i \mod 2$, $m_i=m_{i+1}=m$, and  $|m|\leq \min_j(|\ell_j|)$. Here, $k$ is a negative integer, $\ell$ is the index that labels a representation of $SO(d)$ and $m$ is the index that labels a representation of $SO(d-1)$. 
Before we write the final expression, we highlight two important consequences of our selection rules. Firstly, we can prove that $N$ is always even (i.e. only even powers of $\frac{1}{R}$ occur in the expansion).
\begin{align}
\label{eq:selects}
    \begin{split}
    N &= \sum_{j=1}^q (d-1 +k_j) \mod 2\\
    &= (d-1)n+ \sum_{j=1} (\ell_j-\ell_{j+1}) \mod 2\\
    &= (d-1)n+ \ell_1-\ell_1 \mod 2\\
     &= 0 \mod 2\\
\end{split}
\end{align}
Secondly, this is a finite sum since $\chi_{d,N}$ is a finite set. Therefore, assuming all integrals converge, all sums we write for any term of our expansion for mutual information will be finite.

Recalling that $h(t)=\frac{t}{2}\arccot(t)+\frac{1}{4}\ln(-\frac{1}{4}(t^2+1))$ as introduced in Section \ref{sec:Ent/MI}, we see $h'(-i\coth(\lambda))=i\frac{\lambda}{2}$, which allows us to write the general formula for the $\frac{1}{R^N}$ term of the mutual information in $d$ dimensions. In Section \ref{highSim}, we review spherical harmonic functions. From these we define $\mathcal{N}(d,\ell)$,
\begin{equation}
    \mathcal{N}(d,\ell)= \begin{cases}

        {d+\ell-1 \choose d-1}-{d+\ell-3 \choose d-1} & d\geq 3\\ 1 & d=2  \end{cases},
\end{equation} 
where this is the number of spherical harmonics in spatial dimension $d$ at spherical harmonic degree $\ell$. In the following expression $\mathcal{N}(d-1,m)$ is obtained by summing over all remaining indices in the spherical harmonics, which we labelled by $\eta$,
as described in Sections \ref{highSim} and \ref{spher}. Specifically, Section \ref{spher} uses Schur's Lemma to show that the $\tau$ summation in Equation \eqref{GeneralExapMI} is independent of $\eta$. We take $\mathcal{N}(1,m)$ to be equal to $1$. We have
\begin{equation}\label{GeneralExapMI}
    I_{d,N} =-\sum_{\tau\in\chi_{d,N}}\frac{\mathcal{N}(d-1,m)}{S(\tau)}\left[\prod_{j=1}^{n} \int\diff \lambda_{j}\right]\sum_{q=1}^{n}\left[ \frac{(-1)^{\frac n2}\lambda_{q}\tilde{K}_{\ell_{n},\ell_1,m}^{k_n,d}(\lambda_n,\lambda_{1})\prod_{j=1}^{n-1}\tilde{K}_{\ell_{j},\ell_{j+1},m}^{k_{j},d}(\lambda_{j},\lambda_{j+1})}{\prod_{j=1,j\neq q}^{n} (\coth(\lambda_{q})-\coth(\lambda_{k}))}\right],
\end{equation}
where the expression for $\tilde{K}_{\ell_{j},\ell_{j+1},m}^{k}$ is given by Equations \eqref{Koper} and \eqref{Qfun}, and we remind ourselves that $S(\tau)$ is the number of cyclic permutations that leave $\tau$ unchanged. The full expansion for mutual information in d dimensions is then 
\begin{equation}
    I_d = \sum_{N=0}^\infty I_{d,N} \frac{1}{R^N},
\end{equation}
where we note that if the sum over $\chi_{d,N}$ in Equation \eqref{GeneralExapMI} is empty then the contribution of order $N$ is zero. We will see that the first nonzero term is actually when $N=2d-2$.

\subsection{Mutual Information at Very Large Distances}\label{largeRMI}
We look at the lowest order term in the expansion for Mutual Information. The smallest possible $N$ is $N=2d-2$ which occurs when $\tau = ((0,0,0),(0,0,0))$, $S(\tau)=2$, and $\mathcal{N}(d-1,0)=1$. We can therefore simplify Equation \eqref{GeneralExapMI} dramatically in this case to get
\begin{equation}\label{mi1}
    I_{d,2d-2}= \frac{1}{2}\int_{-\infty}^{\infty}\diff \lambda_1\int_{-\infty}^{\infty}\diff \lambda_2 \text{ } \tilde{K}_{0,0,0}^{0,d}(\lambda_1,\lambda_2) \tilde{K}_{0,0,0}^{0,d}(\lambda_2,\lambda_1) \left[\frac{\lambda_{2}-\lambda_{1}}{\coth(\lambda_{2})-\coth(\lambda_{1})}\right].
\end{equation}
As we can see, we only need to worry about the lowest order term in the expansion of $K(\vectorbold{y},\vectorbold{x})$ in Equation \eqref{kdel},
\begin{equation}\label{orderlow}
    K(\vectorbold{y},\vectorbold{x}) =\frac{\Gamma(\frac{d-1}{2})}{2\pi^{\frac{d+1}{2}}R^{d-1}}
    \begin{bmatrix}
       0&0\\-1&0
    \end{bmatrix} + \mathcal{O}(\frac{1}{R^{d+1}}).
\end{equation}
The corresponding expansion in terms of the spherical wave modes is trivial by noting that the kernel is constant. By noting that the surface area of the $d$-dimensional unit sphere is $\frac{2\pi^{\frac{d}{2}}}{\Gamma(\frac{d}{2})}$, we can write the only nonzero term in the spherical wave mode expansion of $K$:
\begin{equation}
K_{0,0,0}^{0,d}(r,s) =\frac{\Gamma(\frac{d-1}{2})}{\sqrt{\pi}\Gamma(\frac{d}{2})}
    \begin{bmatrix}
       0&0\\-1&0
    \end{bmatrix}.
\end{equation}
To find the kernel with respect to $\alpha_{0,\Omega,d}$, we find $Q^0_{d,0,0}$ using Equation \eqref{Qfun}:
\begin{equation}
     Q^0_{d,0,0}(\lambda)=\frac{\sqrt{\sinh(\lambda)}|\Gamma(i\frac{\lambda}{\pi})|^2|\Gamma(\frac{d-1}{2}+i \frac{\lambda}{\pi})|}{2\sqrt{\lambda} \pi^{\frac32}\Gamma(\frac{d-1}{2})}=\frac{\sqrt{\pi}|\Gamma(\frac{d-1}{2}+i \frac{\lambda}{\pi})|}{2\Gamma(\frac{d-1}{2})\sqrt{\lambda^3\sinh(\lambda)}}.
     \end{equation}
Here, we used the fact that $|\Gamma(ib)|^2=\frac{\pi}{b\sinh(\pi b)}$.
Substituting this into Equation \eqref{Koper} and using the Legendre duplication formula  for the Gamma function, we can find the kernel $\tilde{K}_{0,0,0}^0$, in the $\alpha_0$ basis to get
\begin{equation}\label{ksymMI}
\tilde{K}_{0,0,0}^{0,d}(\lambda_1,\lambda_2) = -\frac{2^{d-2}i}{\pi^2\Gamma(d-1)}\left(\frac{\lambda_2|\Gamma(\frac{d-1}{2}+i\frac{\lambda_1}{\pi})||\Gamma(\frac{d-1}{2}+i\frac{\lambda_2}{\pi})|}{\sqrt{\lambda_1\lambda_2\sinh(\lambda_1)\sinh(\lambda_2)}}\right).
\end{equation}
We substitute this form for $\tilde{K}$ into Equation \eqref{mi1} and find
\begin{equation}
     I \sim -\frac{2^{2d-5}R^{2d-2}}{\pi^4\Gamma(d-1)^2}
     \int_{-\infty}^{\infty}\diff \lambda_1\int_{-\infty}^{\infty}\diff \lambda_2 \text{ } \frac{|\Gamma(\frac{d-1}{2}+i\frac{\lambda_1}{\pi})|^2|\Gamma(\frac{d-1}{2}+i\frac{\lambda_2}{\pi})|^2}{\sinh(\lambda_1)\sinh(\lambda_2)}\left[\frac{\lambda_{2}-\lambda_{1}}{\coth(\lambda_{2})-\coth(\lambda_{1})}\right].
\end{equation}
We now use a hyperbolic identity to rewrite the integrand as
\begin{equation}
     I \sim \frac{2^{2d-5}}{\pi^4\Gamma(d-1)^2R^{2d-2}}
     \int_{-\infty}^{\infty}\diff \lambda_1\int_{-\infty}^{\infty}\diff \lambda_2 \text{ } \left|\Gamma \left(\frac{d-1}{2}+i\frac{\lambda_1}{\pi} \right)\right|^2 \left|\Gamma \left(\frac{d-1}{2}+i\frac{\lambda_2}{\pi} \right) \right|^2\left[\frac{\lambda_{2}-\lambda_{1}}{\sinh(\lambda_{2}-\lambda_{1})}\right]
\end{equation}
Now, we note that $|\Gamma(1+ib)|^2=\frac{\pi b}{\sinh(\pi b)}$ and substitute $u,v = \frac{\lambda_1}{\pi},\frac{\lambda_2}{\pi}$.
\begin{equation}
     I \sim \frac{2^{2d-5}}{\pi^2\Gamma(d-1)^2R^{2d-2}}
     \int_{-\infty}^{\infty}\diff u\int_{-\infty}^{\infty}\diff v \text{ } \left|\Gamma \left(\frac{d-1}{2}+iu \right)\right|^2 \left|\Gamma \left(\frac{d-1}{2}+iv \right)\right|^2|\Gamma(1+i(u-v) )|^2
\end{equation}
The integral over $u$ can be computed with Barnes' Beta Integral \cite{barnes}.\footnote{$\int_{-\infty}^{\infty}\diff s \text{ }\Gamma(a+is)\Gamma(b+is)\Gamma(c-is)\Gamma(d-is) = \frac{2\pi\Gamma(a+c)\Gamma(b+c)\Gamma(a+d)\Gamma(b+d)}{\Gamma(a+b+c+d)}$}
\begin{equation}
     I\sim \frac{2^{2d-4}}{\pi\Gamma(d-1)\Gamma (d+1)R^{2d-2}}
     \int_{-\infty}^{\infty}\diff v \text{ } \left|\Gamma \left(\frac{d+1}{2}+iv \right)\right|^2 \left|\Gamma \left(\frac{d-1}{2}+iv \right) \right|^2
\end{equation}
This integral can also be computed using Barnes' Beta integral and we get the final result.
\begin{equation}
    I \sim \frac{4^{d-1}}{d{2d\choose d}R^{2d-2}} +\mathcal{O}\left(\frac{1}{R^{2d}} \right)
\end{equation}
We can reinsert units and let $r$ be the radius of the two spheres.
\begin{equation}
    I \sim \frac{4^{d-1}r^{2d-2}}{d{2d\choose d}R^{2d-2}} +\mathcal{O} \left(\frac{r^{2d}}{R^{2d}} \right)
    \label{eq:mutinfo}
\end{equation}
As we would expect from the positivity of mutual information, the leading coefficient is positive. 
\begin{table}\centering
\begin{tabular}{ |c|c|c| } 
\hline
Dimension & Exact & Numerical \\\hline 
2& $\frac{1}{3}$& $0.333$ \\ [0.1cm]
3& $\frac{4}{15}$& $0.267$  \\ [0.1cm]
4& $\frac{8}{35}$& $0.229$  \\ [0.1cm]
5& $\frac{64}{315}$& $0.203$  \\ [0.1cm]
6& $\frac{128}{693}$& $0.185$ \\ [0.1cm]
\hline
\end{tabular}\label{MI table}\caption{The leading order coefficient of the mutual information of distant spheres of unit radius for a massless scalar field, obtained using the perturbative method detailed in this paper. These coefficients were obtained previously in \cite{Cardy_2013,Agon2015,Chen_2018} using analytic continuation methods. For comparison to numerical computations as in \cite{Shiba_2012}, we have included the numeric forms of these coefficients.}
\end{table}
For $d=3$, we find the lowest order coefficient is $\frac{4}{15}$. For $d=2$, we get $\frac{1}{3}$. These values match the results of Cardy \cite{Cardy_2013} as well as the extensions by Ag\'on and Faulkner \cite{Agon2015} and Chen et al \cite{Chen_2018} to arbitrary dimensions due to the Legendre duplication formula. 

Furthermore, we note that Equation \eqref{Kgenform} implies that with a suitable solution for the coefficients $F$ bearing indices $k,j,d,$ etc., which amounts to solving a finite dimensional linear algebra problem, we can compute corrections to \eqref{eq:mutinfo} to arbitrary order in $\left(\frac{r^{2d}}{R^{2d}} \right)$. We will comment on this further in the Conclusions \ref{sec:conclusion}. For now, we turn to the computation of the entropy difference for thermal fields.

\section{Entropy Difference of Thermal Massless Fields for $d\geq 3$}\label{TempSect}
In this section, we study fields thermalized at temperature $T$, then restricted to a unit sphere. We will consider the entropy difference between a field at temperature $T$ and the same vacuum field at temperature $0$,
\begin{equation}
   \Delta S(T)= S(\rho_T ) - S(\rho_0).
\end{equation}
In particular, we will construct a series for this quantity for small $T$. The procedure for doing this is very similar to the procedure for the expansion for mutual information. Recalling that $\Delta \alpha_{T,\Omega,d} = \alpha_{T,\Omega,d} - \alpha_{0,\Omega,d}$ is the difference between the $d$-dimensional skewed correlation operators at temperatures $T$ and $0$ on the sphere, we can use the quantum field theoretic version of Equation \eqref{eq:perturbedNice} to write a series for $\Delta S(T)$:
\begin{equation}\label{EntropyDiffExp1}
    \Delta S(T) = \Trace_{L^2(\Omega)^2}\left[h'(\alpha_{0,\Omega,d})\Delta \alpha_{T,\Omega,d}\right]+\sum_{n=2}^\infty\frac{1}{2\pi n i }\oint_\xi \diff z \text{ }h'(z)\Trace_{L^2(\Omega)^2}\left[\left[(z-\alpha_{0,\Omega,d})^{-1}\Delta \alpha_{T,\Omega,d}\right]^{n}\right].
\end{equation}
Recall that the kernel of $\Delta \alpha_{T,\Omega,d}$ is smooth for $d\geq 3$ but has a singularity for $d=2$. Thus, we will only consider $d\geq 3$ since the finiteness of each trace in this expansion is guaranteed by our work in Section \ref{smooth}.

To explicitly compute the series for $ \Delta S(T)$, we 
first compute the expansion for $\Delta\alpha_{T,\Omega,d}$. We can then compute the general expansion for entropy and its lowest order term, just as with mutual information. But, there is an another interesting feature of this expansion. As we saw in Section \ref{popout}, Casini's $K$ corresponds to the first term in the expansion. This is easy to compute in full, which we will do in Section \ref{CasK}.

\subsection{Expanding and Symmetrizing $\Delta \alpha_{T,\Omega,d}$}
Following our derivation for mutual information, we begin by decomposing $\Delta \alpha_{d,\Omega,T}$.
Our goal in this section is to find a series expansion for the difference $\Delta \alpha_{d,\Omega,T}$ for small temperatures assuming $d\geq 3$. 

As discussed in Section \ref{thermal}, finding the kernel of $\Delta \alpha_{d,\Omega,T}$ requires finding the inverse Fourier transform of the functions $f_{d,P}(\vectorbold{k})=|\vectorbold{k}|\left(\coth(\frac{ |\vectorbold{k}|}{2T})-1\right)$ and $f_{d,X}(\vectorbold{k})=\frac{1}{|\vectorbold{k}|}\left(\coth(\frac{ |\vectorbold{k}|}{2T})-1\right)$, where the inverse Fourier transform is taken to be 
\begin{equation}
    g(\vectorbold{x})=\frac{1}{(2\pi)^d}\int_{\mathbb{R}^d} \Diff{d} \vectorbold{k} \text{ }e^{i\vectorbold{k}\cdot\vectorbold{x}}\hat{g}(\vectorbold k).
\end{equation}
We will label these inverse Fourier transforms $g_{d,X}$ and $g_{P,d}$,
We know $g_{d,X}$ and $g_{d,P}$ will be real analytic due to the decay rates of $f_{d,X}$ and $f_{d,P}$. Furthermore, since $f_{d,X}$ and $f_{d,P}$ are spherically symmetric, $g_{d,X}$ and $g_{d,P}$ will be as well. Therefore, $g_{d,X}$ and $g_{d,P}$ are expandable as these radial power series of even order,
\begin{align}
\label{eq:gdxgdp}
    \begin{split}
        g_{d,X}(\vectorbold x) &= \sum_{k=0}^\infty A^k_{d,X} T^{d-1+2k}r^{2k}\\
        g_{d,P}(\vectorbold x) &= \sum_{k=0}^\infty A^{k}_{d,P} T^{d+1+2k}r^{2k}.
    \end{split}
\end{align}
The powers of $T$ are required to balance units. The coefficients $A^k_{d,X}$  and $A^k_{d,P}$ can be found by repeatedly taking the Laplacian and evaluating at zero for a temperature of $T = 1$. To illustrate this, we look at $A^k_{d,X}$, which we can therefore write as
\begin{equation}
    A^k_{d,X}=\frac{(d-2){!}{!}}{(2k){!}{!}(d+2k-2){!}{!}}(\nabla^2)^k g_{d,X}\bigr|_{\vectorbold x=0,T=1}.
\end{equation}
This constant has a convenient representation in Fourier space as
\begin{equation}
   A^k_{d,X} =\frac{(-1)^k(d-2){!}{!}}{(2\pi)^d (2k){!}{!}(d+2k-2){!}{!}}\int_{\mathbb{R}^d}\Diff{d} \vectorbold{k} \text{ } |\vectorbold{k}|^{2k-1}\left(\coth(\frac{ |\vectorbold{k}|}{2})-1\right).
\end{equation}
We can write this integral in radial coordinates as
\begin{equation}
   A^k_{d,X} =\frac{(-1)^{k}(d-2){!}{!}}{2^{d-1}\pi^{\frac{d}{2}}\Gamma(\frac{d}{2}) (2k){!}{!}(d+2k-2){!}{!}}\int_{0}^\infty\diff t \text{ } t^{2k+d-2} \left(\coth(\frac{ t}{2})-1 \right).
\end{equation}
We must now solve the following integral,
\begin{equation}
    I_\mu=\int_0^\infty \diff t \text{ } t^\mu \left(\coth(\frac{t}{2})-1 \right).
\end{equation}
We expand $\coth(\frac{t}{2})-1$ as a power series in $e^{-t}$ and get
\begin{equation}
    I_\mu=2\sum_{n=1}^\infty \int_0^\infty \diff t \text{ } t^\mu e^{-nt}.
\end{equation}
The integral in the sum can be solved with the Gamma function,
\begin{equation}
    I_\mu=2\Gamma(\mu+1)\sum_{n=1}^\infty \frac{1}{n^{\mu+1}}.
\end{equation}
We can now write the sum in terms of the Riemann zeta function, which gives a final expression for this integral as
\begin{equation}\label{cothInt}
    I_\mu= 2\Gamma(\mu+1)\zeta(\mu+1).
\end{equation}
This allows us to write a full expression for $A^k_{d,X}$ and $A^k_{d,P}$ (using identical reasoning) as
\begin{align}\label{aConstforT}
    \begin{split}
   A^k_{d,X} &=\frac{(-1)^k\Gamma(d-1+2k)\zeta(d-1+2k)(d-2){!}{!}}{2^{d-2}\pi^{\frac{d}{2}}\Gamma(\frac{d}{2}) (2k){!}{!}(d+2k-2){!}{!}}\\
   A^k_{d,P} &=\frac{(-1)^{k}\Gamma(d+1+2k)\zeta(d+1+2k)(d-2){!}{!}}{2^{d-2}\pi^{\frac{d}{2}}\Gamma(\frac{d}{2}) (2k){!}{!}(d+2k-2){!}{!}}.
    \end{split}
\end{align}
A comparison with the Taylor series at $T=0$ of the thermal correlation operator for $d=3$ written in Equation \eqref{thermd3} shows that the two match.
These constants allow us to expand the kernel of the correlation matrix for $ \Delta\alpha_{T,\Omega,d}$ in terms of temperature,

\begin{equation}\label{kdelT}
    K_{\Omega,T}^d(\vectorbold{x},\vectorbold{y}) = T^{d-1}\sum_{k=0}^\infty T^{2k}\begin{bmatrix}
        0 & u(k-1)A^{k-1}_{d,P}|\vectorbold x- \vectorbold y|^{2k-2}\\-A^{k}_{d,X}|\vectorbold x- \vectorbold y|^{2k}&0
    \end{bmatrix}.
\end{equation}
Here $u(n)$ is again the Heaviside Step function. Observe that this matrix is is spherically symmetric. Therefore, as discussed in Section \ref{spher}, the only nonzero wave mode blocks are those where the indices of the image and preimage match. Moreover, the nonzero block matrices depend only on $\ell$. Finding these block matrices requires decomposing the power $|\vectorbold x-\vectorbold y|^{2k}$ into spherical harmonics. Similar to the case of mutual information, we note that $|\vectorbold x-\vectorbold y|^{2k}$ is a homogeneous polynomial in both $\vectorbold x$ and $\vectorbold y$ such that the combined degree of in $\vectorbold{x}$ and $\vectorbold{y}$ is $2k$. This means there exists a decomposition into the homogeneous polynomials $G_{\ell,j}^{m,\eta}$ given in Equation \eqref{homBasis}, which we can write as

\begin{align}
    \begin{split}\label{entropyExpand}
         |\vectorbold x-\vectorbold y|^{2k}&= \sum_{\ell=0}^{k}\sum_{j=\ell}^{2k-\ell}\sum_{m,\eta}\bar{F}_{d,\ell,j}^{k}G_{\ell+2j,j}^{m,\eta}(\vectorbold x)^*G_{2k-\ell-2j,k-j-\ell}^{m,\eta}(\vectorbold y)\\&=\sum_{\ell=0}^{k}\sum_{j=\ell}^{2k-\ell}\sum_{m,\eta}\bar{F}_{d,\ell,j}^{k} |\vectorbold x|^{\ell+2j}|\vectorbold y|^{2k-\ell-2j} Y_{\ell}^{m,\eta}\left(\frac{\vectorbold x}{|\vectorbold x|}\right)^* Y_{\ell}^{m,\eta}\left(\frac{\vectorbold y}{|\vectorbold y|}\right).
    \end{split}
\end{align}
This expansion implies an additional selection rule for $\ell$. Namely, $\ell\leq k$. The coefficients $\bar{F}_{\ell,j}^{k}$ can be computed by solving a finite dimensional linear algebra problem over the vector space of polynomials, just as in the case of mutual information. This allows us to write the symmetrized kernel of the $k$-th term of  $\Delta\alpha_{T,\Omega,d}$ in $d$ spatial dimensions, which we will label as $K_{\ell}^{k,d}$, in the form

\begin{equation}\label{KgenformEnt}
K_{\ell}^{k,d}(s,r)=\begin{bmatrix}0&u(k-1)A^{k-1}_{d,P}\sum_{j=\ell}^{2k-\ell-2}\bar{F}_{d,\ell,j}^{k-1}s^{\ell+2j}r^{2k-\ell-2j-2} \\-A^{k}_{d,X}\sum_{j=\ell}^{2k-\ell}\bar{F}_{d,\ell,j}^{k}s^{\ell+2j}r^{2k-\ell-2j}&0
    \end{bmatrix}.
\end{equation}
We note that this has the same functional form as the analogous kernel for mutual information in Equation \eqref{Kgenform}. Exactly as in that case, we can find its kernel with respect to $\alpha_{0,\Omega,d}$, which we label $\tilde{K}_{\ell}^{k,d}$, to be

\begin{align}\label{KoperT}
\begin{split}
   \tilde{K}_{\ell}^{k,d}(\lambda_1,\lambda_2) 
&=-\frac{4i\lambda_2^2\lambda_1}{\pi^2}A^{k}_{d,X}\sum_{j=\ell}^{2k-\ell}\bar{F}_{d,\ell,j}^{k}Q^{\ell+2j}_{d,\ell,0}(\lambda_1)Q^{2k-\ell-2j}_{d,\ell,0}(\lambda_2)\\&-i\lambda_2u(k-1)A^{k-1}_{d,P}\sum_{j=\ell}^{2k-\ell-2}\bar{F}_{d,\ell,j}^{k-1}Q^{\ell+2j}_{d,\ell,0}(\lambda_1)Q^{2k-\ell-2j-2}_{d,\ell,0}(\lambda_2).
\end{split}
\end{align}
Here we reuse the function $Q$ defined in in Equation \eqref{Qfun} and we remind the reader that $d$ is the spatial dimension, that $\ell$ denotes the $SO(d)$ index of the wave mode subspace which is guaranteed by Schur's Lemma to be the same for the image and preimage and that $k$ means that we are looking at the order $T^k$ term of $\Delta\alpha_{T,\Omega,d}$.

\subsection{General Expansion for the Entropy Difference and Relative Entropy}\label{genExpandEnt}
We can now describe the general form of the $T^N$ term of the expansion for the entropy difference. As we showed in Section \ref{popout}, the entropy difference can be split into the relative entropy and Casini's $K$. Assume that Casini's $K$ can be decomposed into a power series $K_\Omega(T)=\sum_{N=1}^\infty Q_{d,N}T^N$, where we label Casini's $K$ by $K_\Omega(T)$. As we will see in Section \ref{CasK}, computing these coefficients is relatively straightforward because $h'(\alpha_{0,\Omega,d})$ is a differential operator. So, we focus on the decomposition of relative entropy.

We can use the results of Section \ref{petuebSphere}. Just as with mutual information, our selection rules for $\ell,m,\eta$ will limit to finitely many paths,  meaning all sums we write for any term of our expansion for mutual information will be finite. We parameterize the terms as follows. Given $d$ and $N$, define $\bar{\chi}_{d,N}$ to be the set of all circular lists of tuples of the form $(k,\ell)$ such that $\sum_{j=1}^n (d-1 + 2k_j) = N$, $n\geq 2$, $\ell_j=\ell_{j+1}=\ell$, and $\ell\leq \min_j(k_j)$. Here, $k$ and $\ell$ are non-negative integers. 

Recalling that that $h'(-i\coth(\lambda))=i\frac{\lambda}{2}$, we write the general formula for the $T^N$ term of the relative entropy in $d$ dimensions as
\begin{equation}\label{GeneralExapEnt}
    \bar{S}_{d,N} =-\sum_{\tau\in\bar\chi_{d,N}}\frac{\mathcal{N}(d,\ell)}{2S(\tau)}\left[\prod_{j=1}^{q} \int\diff \lambda_{j}\right]\sum_{q=1}^{n}\left[ \frac{i^n\lambda_{q}\tilde{K}_{\ell}^{k_n,d}(\lambda_n,\lambda_{1})\prod_{j=1}^{n-1}\tilde{K}_{\ell}^{k_{j},d}(\lambda_{j},\lambda_{j+1})}{\prod_{j=1,j\neq q}^{n} (\coth(\lambda_{q})-\coth(\lambda_{k}))}\right].
\end{equation}
Note that the $\mathcal{N}(d,\ell)$ comes from summing over all $m,\eta$ since the summand is independent of it. We can then find the coefficients of the entropy difference using $S_{d,N}=Q_{d,N}-\bar{S}_{d,N}$. Given these coefficients, the relative entropy and the entropy difference have the following expansions:
\begin{equation}
    S(T||0) = \sum_{N=1}^{\infty}\bar{S}_{d,N}T^{N}
\end{equation}
\begin{equation}
    S(T)-S(0) = \sum_{N=1}^{\infty}S_{d,N}T^{N}.
\end{equation}

\subsection{Evaluating Casini's $K$}\label{CasK}

Casini's $K$ is straightforward to compute using the spatial basis. Using the form of the modular Hamiltonian in Equation \eqref{modHamSphere} and recalling the definitions of $g_{d,X}$ and $g_{d,P}$ as the matrix entries of the kernel of $\Delta\alpha_{T,\Omega,d}$ (see text above Equation \eqref{eq:gdxgdp}), we find the kernel of $\frac{1}{2}h'(\alpha_{0,\Omega,d})\Delta\alpha_{T,\Omega,d}$ to be

\begin{equation}
    K(\vectorbold x,\vectorbold y) = \frac\pi4
    \begin{bmatrix}
        -\nabla_{\vectorbold x}\cdot(p(\vectorbold x)\nabla_{\vectorbold x}g_{d,X}(\vectorbold{x}-\vectorbold{y}))+(d-1)g_{d,X}(\vectorbold{x}-\vectorbold{y})&0\\0&p(\vectorbold x) g_{d,P}(\vectorbold{x}-\vectorbold{y})
    \end{bmatrix}.
\end{equation}
To find the trace, we take the trace of this $2\times 2$ matrix, set $\vectorbold y = \vectorbold x$ and integrate over $\Omega$:
\begin{equation}
    K_\Omega(T) =\frac\pi4\int_{\Omega} \diff^d \vectorbold{x} \text{ }(d-1)g_{d,X}(0)+p(\vectorbold x)( g_{d,P}(0) -\nabla^2g_{d,X}(0)).
    \label{eq:kernelt}
\end{equation}
 We can solve the remaining integrals in Equation \eqref{eq:kernelt} by using the formulae for the volume and surface area of the unit sphere,
\begin{equation}
     K_\Omega(T) = \frac{\pi^{\frac{d}{2}+1}}{4\Gamma(\frac {d} {2}+1)} ((d-1)g_{d,X}(0)+ \frac{2}{d+2} (g_{d,P}(0)-\nabla^2g_{d,X}(0))).
\end{equation}
Now using Equation \eqref{aConstforT}, we can entirely solve for $ K_\Omega(T)$:
\begin{equation}
     K_\Omega(T) = \frac{\sqrt{\pi}\Gamma(\frac{d+1}{2})}{2\Gamma(\frac{d}{2}+1)}\left(\zeta(d-1)T^{d-1}+\frac{4d\zeta(d+1)}{d+2}T^{d+1}\right).
\end{equation}
We see that $K_\Omega(T)$ contributes terms of order $T^{d-1}$ and $T^{d+1}$ to the entropy difference.

\subsection{Entropy Difference for Very Small Temperatures}
We look at the lowest order term in the expansion for relative entropy, cf.~Equation \ref{eq:peturbRelEnt}. The smallest possible $N$ is $N=2d-2$ which occurs when $\tau = ((0,0),(0,0))$, $n=2$, $S(\tau)=2$, and $\mathcal{N}(d,0)=1$. We find an expression for the leading coefficient that is similar to the coefficient for mutual information,

\begin{equation}\label{E1}
    S(T||0)\sim T^{2d-2}\frac{1}{4}\int_{-\infty}^{\infty}\diff \lambda_1\int_{-\infty}^{\infty}\diff \lambda_2 \text{ } \tilde{K}_{0}^{0,d}(\lambda_1,\lambda_2) \tilde{K}_{0}^{0,d}(\lambda_2,\lambda_1) \left[\frac{\lambda_{2}-\lambda_{1}}{\coth(\lambda_{2})-\coth(\lambda_{1})}\right].
\end{equation}
We again only need to worry about the lowest order term in the expansion of $K_{d,\Omega,T}(\vectorbold{y},\vectorbold{x})$ in Equation \eqref{kdelT},
\begin{equation}\label{orderlowT}
    K(\vectorbold{y},\vectorbold{x}) =\frac{\Gamma(\frac{d-1}{2})\zeta(d-1)T^{d-1}}{\pi^{\frac{d+1}{2}}}
    \begin{bmatrix}
       0&0\\-1&0
    \end{bmatrix} + \mathcal{O}(T^{d+1}).
\end{equation}
We note that this has the same form as the lowest order term in Equation \eqref{orderlow} up to a factor of $2\zeta(d-1)$. Because of this, and the fact that the form of the Equation \eqref{E1} is the same as Equation \eqref{mi1}, we can use our result for the leading coefficient of mutual information to obtain
\begin{equation}
     S(T||0)\sim 4\zeta(d-1)^2I_{d,2d-2}T^{2d-2}=\frac{2^{2d-1}\zeta(d-1)^2}{d{2d\choose d}}T^{2d-2}.
\end{equation}

As we would expect from the positivity of relative entropy, its leading term is positive. This preserves the inequality $\Delta S(T) \leq K_\Omega(T)$ to lowest order. We now have the first three leading coefficients of the entropy difference:
\begin{equation}
    \Delta S(T) \sim   \frac{\sqrt{\pi}\Gamma(\frac{d+1}{2})}{2\Gamma(\frac{d}{2}+1)}\left(\zeta(d-1)T^{d-1}+\frac{4d\zeta(d+1)}{d+2}T^{d+1}\right)-\frac{2^{2d-1}\zeta(d-1)^2}{d{2d\choose d}}T^{2d-2} +\mathcal{O}(T^{2d}).
\end{equation}
Note that if we reinsert units and let $r$ be the the radius of the sphere, we get
\begin{equation}
    \Delta S(T) \sim  \frac{\sqrt{\pi}\Gamma(\frac{d+1}{2})}{2\Gamma(\frac{d}{2}+1)}\left(\zeta(d-1)(rT)^{d-1}+\frac{4d\zeta(d+1)}{d+2}(rT)^{d+1}\right)-\frac{2^{2d-1}\zeta(d-1)^2}{d{2d\choose d}}(rT)^{2d-2} +\mathcal{O}((rT)^{2d}).
\end{equation}
The lowest order term is therefore proportional to the surface area of the sphere. 
\begin{table}\label{enttable}\centering
\begin{tabular}{ |c|c|c| } 
\hline
Dimension & Exact & Numerical \\\hline 
3& $\frac{\pi^2}{9}(rT)^2+\frac{2\pi^4}{675}(rT)^4$& $1.097(rT)^2+0.289(rT)^4$ \\ [0.1cm]
4& $\frac{3\pi \zeta(3)}{16}(rT)^3+\frac{\pi \zeta(5)}{2}(rT)^5-\frac{16 \zeta(3)^2}{35}(rT)^6$& $0.708(rT)^3+1.629(rT)^5-0.661 (rT)^6$  \\ [0.1cm]
5& $\frac{4\pi^4}{675}(rT)^4+\frac{32\pi^6}{19845}(rT)^6-\frac{32\pi^8}{637875}(rT)^8$& $0.577(rT)^4+1.550(rT)^6-0.476 (rT)^8$  \\ [0.1cm]
6& $\frac{5\pi \zeta(5)}{32}(rT)^5+\frac{15\pi \zeta(7)}{32}(rT)^7-\frac{256 \zeta(5)^2}{693}(rT)^{10}$& $0.509(rT)^5+1.485(rT)^7-0.397 (rT)^{10}$ \\ [0.1cm]
\hline
\end{tabular}\caption{The leading order coefficients of entropy difference between a low temperature field and the vacuum on a sphere for a massless scalar field in exact and numerical form. Note that the terms in the expansion are proportional to $rT$ for a sphere of radius $r$ and temperature $T$.}
\end{table}

\section{Entropy Difference and Area Laws}\label{AreaSect}
We observed that for a thermal field, the lowest order term of the entropy expansion had an `area law' term. In this Section, we will show that this is a general feature of low entropy expansions for entropy difference on the sphere so long as the kernel of $\Delta\alpha$ is smooth.

Consider a skewed correlation matrix family $\alpha(t_1,\dots,t_q)$ with $\alpha(0,\dots,0)=\alpha_{0,\Omega,d}$ where the $t_i$ have units of energy. Suppose we have a general expansion for $\Delta\alpha=\alpha(t_1,\dots,t_q)-\alpha(0,\dots,0)$,
\begin{equation}
    \Delta\alpha= \sum_{N=1}^\infty\left[\sum_{\sum k_i=N}\alpha_{k_1,\dots,k_q}t_1^{k_1}\dots t_q^{k_q}\right].
\end{equation}
Consider the corresponding expansion of the kernel of $  \Delta\alpha$:
\begin{equation}
   K(\vectorbold x,\vectorbold y)= \sum_{N=1}^\infty\left[\sum_{\sum k_i=N}K_{k_1,\dots,k_q} (\vectorbold x,\vectorbold y)t_1^{k_1}\dots t_q^{k_q}\right].
\end{equation}
We note that the entries of $K(\vectorbold x,\vectorbold y)$ will have different units following from their definitions. Reminding ourselves of the form of the entries of the skewed correlation matrix,
\begin{equation}
  \alpha =  \begin{bmatrix}
            V_{off}^T&P\\-X&-V_{off}
        \end{bmatrix},
\end{equation}
we see that the kernel for the $X$ entry must have units of  $E^{d-1}$ where $E$ is energy, the kernel of the $P$ entry must have units of  $E^{d+1}$, and the kernel of the $V_{off}$ and $V_{off}^\dagger$ entries must have units of $E^{d}$. Since the scalar field is conformal, the only quantities with units in the problem are $\vectorbold x$, $\vectorbold y$, and the $t_i$. So by unit analysis, if an entry of $K(\vectorbold x,\vectorbold y)$ has units of $\delta$, the corresponding entry of $K_{k_1,\dots,k_q}(\vectorbold x,\vectorbold y)$ must be a homogeneous function of order $\sum_i k_i-\delta$. But we also know that the entries of $K$ are smooth, meaning each of the entries of the $K_{k_1,\dots,k_q}$ are smooth. Since we cannot have a smooth homogeneous function of an order less than zero, the lowest order term for each entry must have order $\delta$. The lowest order term for $K$ must then be of order $\min(d,d-1,d+1)=d-1$ with the only nonzero entry being a constant function in the $X$ component,
\begin{equation}
   K(\vectorbold x,\vectorbold y)\sim \sum_{\sum k_i=d-1}\begin{bmatrix}
       0&0\\-B_{k_1,\dots,k_q} &0
   \end{bmatrix}t_1^{k_1}\dots t_q^{k_q}.
\end{equation}
Now consider the lowest order term for the entropy measured on a unit sphere, making a choice of units such that $R=1$. Just as with the thermal field, it will originate from Casini's $K$ and we get 
\begin{equation}
    \Delta S\sim \sum_{\sum k_i=d-1}t_1^{k_1}\dots t_q^{k_q}\Trace_{L^2(\Omega)^2}\left[h'(\alpha_{0,\Omega,d})\begin{bmatrix}
       0&0\\-B_{k_1,\dots,k_q} &0
   \end{bmatrix}\right].
\end{equation}
But, we can easily compute this using the same procedure as Section \ref{CasK}. If $A_d$ is the surface area of the unit sphere, we have 
\begin{equation}
    \Delta S\sim \frac{\pi(d-1)}{4d}\left[\sum_{\sum k_i=d-1}B_{k_1,\dots,k_q}t_1^{k_1}\dots t_q^{k_q} \right]A_d.
\end{equation}
But now we can reinsert units to consider a sphere of any radius $r$,
\begin{equation}
    \Delta S\sim \frac{\pi(d-1)}{4d}\left[\sum_{\sum k_i=d-1}B_{k_1,\dots,k_q}t_1^{k_1}\dots t_q^{k_q} \right](A_dr^{d-1}).
\end{equation}
So we do indeed see an area law for the entropy difference in lowest order.

\subsection{On the Universality of the Area Law}\label{AreaLawCaveats}
A natural question is to ask whether this area law can be generalized to an arbitrary quantum field theory. We  briefly comment here.

First, we note that the area law is the most general lowest order term, but that this term may be zero in some cases. If so, we expect the next lowest order term to originate from $V_{off}$ and be of order $r^d$. However, if the cases studied here are any indication, we expect $V_{off}$ to often be zero, producing no contribution as well. In this case, the next lowest order term is originates from the momentum correlation function $P$ and is of order $r^{d+1}$. In particular, if the perturbation is smooth and the perturbed state is also conformal, the power law for position correlation function of conformal field theories implies that $X$ is unchanged. This means the area law term is zero and we would expect the lowest order term to be of order $r^d$ or $r^{d+1}$. For example, Blanco et al \cite{Blanco:2013joa} perturbs a scalar field on a sphere in $D-2$ spatial dimensions into a general holographic conformal state and find the entropy difference is of order $r^{D}$.

Second, we note that order of the lowest order term comes from the dimensions of $X$ which itself ultimately arises from the fact that the scaling dimension of a massless scalar field in $d$ spatial dimensions is $\Delta = \frac{d-1}{2}$. In general, when perturbing from an arbitrary conformal field theory with scaling dimension $\Delta$, the most general lowest order term would be of order $r^{2\Delta}$. Since the scaling dimension is in general larger than $\frac{d-1}{2}$, we would not expect to see an area law when perturbing from an arbitrary conformal field theory.

\section{Conclusion}
\label{sec:conclusion}

We have outlined a new perturbative strategy for computing an expansion for the entropy and mutual information of nearly conformal states on a sphere. We applied this to the problem of computing the mutual information of two distant spheres in a massless scalar field and found that the lowest order term agrees with preexisting literature. We then applied this to the case of the entropy difference of thermal fields and found the lowest order terms. In particular, we found that the lowest order term is proportional to the area of the sphere, which we showed was a general feature of a low energy expansion for the energy difference.

Given that we have a full expansion for the mutual information and the entropy difference, it would be natural to use them to compute the next to leading order term and compare them with other results in the literature such as Chen et al \cite{Chen_2017}. It would also be valuable to look at a thermal field for the $d=2$ case. Recalling that the correlation matrix $X$ has an unremovable logarithmic divergence, we note that Casini's K is clearly infinite using the results of Section \ref{CasK}. However, we note that the relative entropy may still be finite, as the traces appearing in its expansion are finite, owing to the fact that $\Delta\alpha(T)$ is Hilbert Schmidt for $d=2$. It may be the case that the entropy is infinite but the relative entropy is finite.

Additionally, the approach we have outlined is very general and extends to any physical quantity which is a matrix function of the skewed correlation matrix. These include the R{\'{e}}nyi entropy, the standard R{\'{e}}nyi mutual information used by Cardy to analytically continue to the standard mutual information, the Petz-Rényi relative entropy between a state and the vacuum \cite{Seshadreesan_2018}, and the alternative definition of R{\'{e}}nyi mutual information given by Kudler-Flam \cite{AltRenyi}. In particular, it would be interesting to compute the standard R{\'{e}}nyi mutual information for integer $n$ and compare with Cardy. The perturbative approach we have laid out is particularly valuable because, even if the coefficients of an expansion do not have a simple analytic form, one can still numerically integrate to compute them. Since the number of iterated integrals needed to compute the coefficients depends only of the order of the coefficients and not on the dimension of the underlying field theory, this approach provides a numerical strategy for computing these expansions which avoids some troubles of dimensionality present in lattice field theory. 

Furthermore, our approach should extend to general Gaussian bosonic states. In particular, future work could look at both the mutual information and the entropy difference for a thermal \textit{massive} scalar field by expanding the correlation matrix in terms of mass as well as distance and/or temperature. We also expect there to be an analogous strategy for expansions of nearly conformal fermionic Gaussian states built from the continuum limit of fermionic lattice field theory.

On the mathematical side of things, the derivation of this expansion presented in this work is at times non-rigorous and more work could be done to establish the validity of both the general expansion for an analytic function of a linear mapping in Section \ref{generalexpan} and the specific expansion for Gaussian states on a sphere presented in Section \ref{petuebSphere}. We made some progress in Section \ref{smooth} by explaining why the traces in each term were finite. However, we did not establish why the integrals for each term were finite other than by direct computation.

We also point out that our derivation for the perturbation of a function of the skewed correlation matrix relies only on the fact that the set of skewed correlation matrices is a convex cone whose spectra lie in a specific subset of the complex plane. We note that this property is shared by the set of  density operators. Therefore, the mathematical approach of perturbing the resolvent may also give expansions for the entropy difference and mutual information of nearly conformal non-Gaussian states when applied to the  density operator rather than the skewed correlation matrix. A particular simple case to study would be a weakly interacting $\lambda \phi^4$ theory.

Finally, we would be remiss not to point out the connection between quantum field theory and gravity implied by the appearance of an area law in lowest order for the entropy difference. It would be interesting to investigate any connection between this area law and the numerical one observed by Srednicki. Perhaps the entropy of lattice field theory with lattice separation $a$ is related to the entropy difference between the scalar vacuum on a sphere and some state with energy scale $a^{-1}$.

It would also be valuable to investigate to what extent the area law for entropy difference is a general characteristic of a quantum field theory for a smoothly perturbed state, bearing in mind the caveats pointed out in Section \ref{AreaLawCaveats}. It may be that the origin of the black hole entropy would be as the lowest order term in the expansion of the entropy difference in some quantum theory of gravity. This would be akin to the classical kinetic energy difference appearing as the lowest order term in the expansion for the relativistic kinetic energy difference. If true, the black hole entropy would really be an entropy difference $
    \Delta S_{\text{BH}} = \frac{\Delta A}{4G}$.
This is of course pure speculation. Indeed, the existence of an area law for a scalar field theory may simply be a coincidence arising from the fact that the scaling dimension of a scalar field theory happens to be $\frac{d-1}2$, giving the two point correlation function the same units as inverse area.
\appendix

\section{Gaussian States in Lattice Field Theory} 
\label{sec:gaussLattice}
In this Appendix, we provide a general discussion of Gaussian states. Gaussian states are a special kind of state which are completely described by the two point correlation functions. All states that we care about are Gaussian. To avoid mathematical technicalities, let us only consider systems with finite dimensional configuration spaces. The first three sections give the definition of and an example of a Gaussian state. The next section discusses the decomposition of Gaussian states into a tensor product of one dimensional harmonic oscillator states. We then use this decomposition to find formulas for entropy and mutual information in terms of traces in the configuration space. Finally, we describe how to perturb the entropy and mutual information of a state whose Gaussian decomposition is known to a state where the decomposition isn't known. This perturbative technique is the strategy we use to compute the central results of this paper.

\subsection{Wigner Functions}\label{Wigner}
To discuss Gaussian states, we first need to introduce the concept of Wigner functions. Suppose that we have a quantum system with a configuration space $C$.  The Wigner function is a function of the vector $\vectorbold{v} = (\vectorbold{v}_1,\vectorbold{v}_2)\in C^2$ and holds all of the same information as the density operator, completely describing a quantum state. It can be obtained from the density operator $\hat\rho$,
\begin{equation}
   W(\vectorbold{v}) = \frac{1}{\pi^n}\int \diff \vectorbold{q} \text{ }e^{2i\vectorbold{v}_2\cdot\vectorbold{q}}\langle \vectorbold{v}_1-\vectorbold{q}|\hat{\rho}|\vectorbold{v}_1+\vectorbold{q}\rangle.
\end{equation}
Given $\vectorbold{w} \in C$ with components $w_k$, the ket $|\vectorbold{w}\rangle$ denotes the simultaneous eigenvector of the field operators such that  $\hat{\phi}_k|\vectorbold{w}\rangle=w_k|\vectorbold{w}\rangle$.
\subsection{Gaussian States}
In general, computing thermodynamic or quantum informational quantities for arbitrary quantum states can be difficult. To simplify calculations, we will only consider a subset of states known as Gaussian states, which are completely determined by their two point correlation functions. A Gaussian state is defined as a state with Wigner function,
\begin{equation}
   W(\vectorbold{v}) = \frac{1}{\sqrt{\det(\pi\sigma)}}\exp(-(\vectorbold{v}-\vectorbold{v}_0)^\dagger\sigma^{-1}(\vectorbold{v}-\vectorbold{v}_0)).
\end{equation}
A general review of Gaussian states is given by Ferraro~et~al. \cite{ferraro2005gaussian}, from which much of this section is borrowed.
Here $\sigma$ is the \textit{covariance matrix}, which is a real symmetric matrix and $\vectorbold{v}_0$ is the mean vector. They are defined by the following equations,
\begin{align}
    \sigma_{jk} &=\langle \{\hat{V}_j,\hat{V}_k\}\rangle-2 \langle \hat{V}_j\rangle\langle \hat{V}_k\rangle,
    &    \vectorbold{v}_{0,j} &= \langle \hat{V}_j\rangle,
\end{align}
where $\{\cdot,\cdot\}$ denotes an anti-commutator, $\langle \cdot \rangle$ denotes an expectation value and $\hat{V}_j$ denotes the components of he vector valued operator which combines the fields and their conjugate momenta: $\hat{\vectorbold{V}}=(\hat{\boldsymbol{\phi}},\hat{\boldsymbol{\pi}})$.
From the canonical commutation relation (see Appendix \ref{app:math}, Equation \eqref{uncertainty}) $\sigma$ must be such that $\sigma+iJ$ is positive definite.
Here, we represent by $J$ the matrix
\begin{equation}
    J =\begin{bmatrix}
        0& I\\-I&0\\
    \end{bmatrix},
\end{equation} where $I$ is the identity matrix on $C$. Because $\sigma$ is real, this means that the complex conjugate of  $\sigma+iJ$, which is $\sigma-iJ$, is also positive definite. Therefore, we can show
\begin{equation}
    \sigma = \frac{1}{2}((\sigma+iJ)+(\sigma-iJ)) > 0, 
\end{equation}
meaning $\sigma$ itself is also positive definite. Note that knowing the mean vector and the covariance matrix is equivalent to knowing the state. In every example that we will look at in this paper, the mean vector will be zero,
\begin{equation}
    \vectorbold{v}_{0,j}=\langle \hat{V}_j\rangle = 0.
\end{equation}
We will call such Gaussian states \textit{centered}. Centered Gaussian states are therefore completely determined by their covariance matrix. Additionally, it will sometimes be convenient to split $\sigma$ into four square block matrices for our purposes,
\begin{equation}\label{SplitCorr}
    \sigma = \begin{bmatrix}
        X & V_{off}\\ V_{off}^T & P\\
    \end{bmatrix},
\end{equation}
here $X$, $P$, and $V_{off}$ are the matrices with entries given by the two point correlation functions defined in Equation \eqref{FiniteCor}.

A useful fact about Gaussian states is that partial traces of Gaussian states are also Gaussian states. To be precise, let $H_1$ be a subsystem of $H$ with configuration spaces $C_1$ and $C$ respectively. If $\rho$ is a Gaussian state in $H$ with correlation matrix $\sigma$, then $\rho_{C_1}$ is Gaussian in $H_1$ with correlation matrix equal to the block matrix $\sigma_{C_1,C_1}$, as defined in Section \ref{block}. 

\subsection{Generalized Thermal Coupled Oscillator}\label{genThermalOscillator}
An important case of a Gaussian state is the \textit{generalized thermal coupled oscillator}. Consider $n$ coupled oscillators with Hamiltonian $2\hat H = \sum_{i=1}^n\hat \pi_i^2 + \sum_{ij}K_{ij} \hat{\phi}_i\hat{\phi}_j$ where $K$ is an $n \times n$ positive definite coupling matrix. 

Now, consider a Bogoliubov transformation from the standard vector valued field operators $\hat{\boldsymbol\phi}$ and $\hat{\boldsymbol\pi}$ to normal modes $\hat{\boldsymbol\phi}'=Q^T\hat{\boldsymbol\phi}$ and $\hat{\boldsymbol\pi}'=Q^T\hat{\boldsymbol\pi}$. Here $Q$ is the matrix whose columns are the eigenvectors of $K$.  This transforms the Hamiltonian into a commuting sum of one-particle Hamiltonians with normal frequencies given by $\omega_i$, the square rooted eigenvalues of $K$, 
\begin{equation}
   2\hat H = 2\sum_{i=1}^n \hat H_i=\sum_{i=1}^n  (\hat{\pi_i'}^2 + \omega_i^2  \hat{\phi_i'}^2).
\end{equation}
We can construct a Gaussian state by joining each individual normal mode of the oscillator with frequency $\omega_i$ to a heat bath with inverse temperature $\beta_i$. This is described by the density matrix
\begin{equation}
\hat{\rho} = 2^n\bigotimes_{i=1}^n \sinh(\frac{\beta_i\omega_i}{2})\exp(-\beta_i\hat{H_i}),
\end{equation}
where $\hat{\phi_i'}$ and $\hat{\pi_i'}$ are the operator components of $\hat{\boldsymbol \phi}'$ and $\hat{\boldsymbol \pi}'$. Since a one particle thermal oscillator is Gaussian, the entire tensor product state will also be Gaussian. Its two point correlation functions in terms of the normal modes operators can be easily computed as,
\begin{align}
2\langle \hat{\phi'_j}\hat{\phi'_k}\rangle&=\frac{1}{\omega_j}\coth(\frac{\beta_j \omega_j}{2}) \delta_{jk} &  2\langle \hat{\pi'_j}\hat{\pi'_k}\rangle  &=\omega_j\coth(\frac{\beta_j \omega_j}{2}) \delta_{jk}&\langle\{\hat{\phi'_j},\hat{\pi'_k}\}\rangle &=0.
\end{align}

Now we can invert the Bogoliubov transformation and write the correlation matrices of the original field operators as
\begin{align}\label{GenThermHam}
    X =2\langle \hat{\phi}_j\hat{\phi}_k\rangle&=  \frac{1}{\sqrt{K}}\coth(\frac{B \sqrt{K}}{2}), &P=2\langle \hat{\pi}_j\hat{\pi}_k\rangle &=\sqrt{K}\coth(\frac{B \sqrt{K}}{2}), & V_{off}=\langle\{\hat{\phi_j},\hat{\pi}_k\}\rangle &= 0.
\end{align}
Here we define a symmetric matrix $B$ which has the same eigenvectors as $K$ but with eigenvalues $\beta_i$.
\subsection{Decomposing a Arbitrary Gaussian State}
 We start with the Wigner function of a centered Gaussian state,
\begin{equation}
   W(\vectorbold{v}) = \frac{1}{\sqrt{\det(\pi\sigma)}}\exp(-\vectorbold{v}^\dagger\sigma^{-1}\vectorbold{v}),
\end{equation}
where $\sigma$ is the correlation matrix.
From the Takagi Transformation reviewed in Section \ref{Takagi}, there exists a symplectic matrix $S$ such that $S^T\sigma S=D$ where $D_0=\mathrm{diag}(\lambda_1,\dots,\lambda_n)$ with each $\lambda_i>1$ and  \begin{equation}\label{diagonalCorr}
    D=
    \begin{bmatrix}
    D_0&0\\0&D_0
\end{bmatrix}.
\end{equation}
We can rewrite this expression as $\sigma^{-1}=SD^{-1}S^T$. Therefore if we canonically transform the vector $\vectorbold{v}$ into $\vectorbold{w}=S^T\vectorbold{v}$, we can transform an arbitrary Gaussian state into one with diagonal Wigner function,
\begin{equation}
   W(\vectorbold{w}) = \frac{1}{\sqrt{\det(\pi D)}}\exp(-\vectorbold{w}^\dagger D^{-1}\vectorbold{w}).
\end{equation}
That is, this is now the Wigner function of a Gaussian state with a diagonal correlation matrix $D$. Comparing Equations \eqref{diagonalCorr}, \eqref{SplitCorr}, and \eqref{GenThermHam}, we see that this new Gaussian state is equivalent to a generalized thermal coupled oscillator with $B=2\mathrm{arccoth}(D)$ and $K=I$. In other words, we can decompose an arbitrary Gaussian state as the tensor product of $n$ independent oscillators all with unit frequency at inverse temperatures $\beta_i=2\mathrm{arccoth}(\lambda_i)$, noting that $\lambda_i > 1$.

We wish to explicitly write the decomposed  density operator of our Gaussian state. To do so, we must first define the transformed vector valued field operators $\hat{\boldsymbol\phi}'$ and $\hat{\boldsymbol\pi}'$ corresponding to $\vectorbold w$ as 

\begin{equation}
    \begin{bmatrix}
        \hat{\boldsymbol\phi}' \\ \hat{\boldsymbol\pi}'
    \end{bmatrix} = S^T \begin{bmatrix}
        \hat{\boldsymbol\phi} \\ \hat{\boldsymbol\pi}
    \end{bmatrix}.
\end{equation}

Our density operator can now be written as the tensor product of $n$ canonical ensembles with unit frequency, at inverse temperatures $\beta_i=2\mathrm{arccoth}(\lambda_i)$,

\begin{equation}\label{DecompGauss}
    \hat{\rho} = \bigotimes_{i=1}^n \frac{2}{\sqrt{\lambda_i^2-1}}\exp(-\mathrm{arccoth}(\lambda_i)( \hat{\phi_i'}^2+ \hat{\pi_i'}^2)),
\end{equation}
where $\hat{\phi_i'}$ and $\hat{\pi_i'}$ are the operator components of $\hat{\boldsymbol \phi}'$ and $\hat{\boldsymbol \pi}'$.
\subsection{The Modular Hamiltonian}
The general decomposition given above decomposition allows us to write the operator logarithm of $\hat \rho$,
\begin{equation}
    \ln(\hat{\rho}) = -\sum_{i=1}^n \left[\mathrm{arccoth}(\lambda_i)(\hat{X}_{W,i}^2+\hat{P}_{W,i}^2)+\frac{1}{2}\ln(\frac{1}{4}(\lambda_i^2-1))\right].
\end{equation}
We can turn this back into matrix form using a trace over the space $C^2$ by using the vector operator $\hat{\vectorbold{W}} = (\hat{\boldsymbol \phi}',\hat{\boldsymbol \pi}')$, defined analogously to the vector valued operator $\hat{\vectorbold{V}}$. The operator logarithm becomes,
\begin{equation}
    \ln(\hat{\rho}) = -\Trace_{C^2} \left[\hat{\vectorbold{W}}^{T}\begin{bmatrix}
        \mathrm{arccoth}(D_0)&0\\0& \mathrm{arccoth}(D_0)
    \end{bmatrix}\hat{\vectorbold{W}}+\begin{bmatrix}
       \frac{1}{4}\ln(\frac{1}{4}(D_0^2-1))&0\\0& \frac{1}{4}\ln(\frac{1}{4}(D_0^2-1))
    \end{bmatrix}\hat{I}\right].
\end{equation}
Note that the transpose here denotes a transpose in the configuration space. Now, we invert the Bogoliubov transformation and write $\hat{\vectorbold{W}}=S^T\hat{\vectorbold{V}}$,
\begin{equation}\label{LogOperatorBad}
    \ln(\hat{\rho}) = -\Trace_{C^2} \left[\hat{\vectorbold{V}}^TS\begin{bmatrix}
        \mathrm{arccoth}(D_0)&0\\0& \mathrm{arccoth}(D_0)
    \end{bmatrix}S^T\hat{\vectorbold{V}}+\begin{bmatrix}
       \frac{1}{4}\ln(\frac{1}{4}(D_0^2-1))&0\\0& \frac{1}{4}\ln(\frac{1}{4}(D_0^2-1))
    \end{bmatrix}\hat{I}\right].
\end{equation}
We can rewrite the first matrix in a more convenient form (see Appendix \ref{app:math}, Equation \eqref{similiar}) and defining $f(t)=\arccot(t)$ for brevity,
\begin{align}\label{SymLogRho}
\begin{split}
    S\begin{bmatrix}
        \mathrm{arccoth}(D_0)&0\\0& \mathrm{arccoth}(D_0)
    \end{bmatrix}S^T&=-S\begin{bmatrix}
        0&\mathrm{arccoth}(D_0)\\-\mathrm{arccoth}(D_0)& 0
    \end{bmatrix}JS^T\\&=-S\begin{bmatrix}
        0&\mathrm{arccoth}(D_0)\\-\mathrm{arccoth}(D_0)& 0
    \end{bmatrix}S^{-1}J\\&=S\arccot(\begin{bmatrix}
        0&D_0\\-D_0& 0
    \end{bmatrix})S^{-1}J
    \\&=f(J\sigma)J.
    \end{split}
\end{align}
Here, we use the matrix $J$ defined in Equation \eqref{eq:J} in Appendix \ref{sec:symp}. Now defining the function $g(t) = \frac{1}{4}\ln(-\frac{1}{4}(t^2+1))$, we can also rewrite the second matrix,
\begin{align}
    \begin{split}
    \begin{bmatrix}
        \frac{1}{4}\ln(\frac{1}{4}(D^2-1))&0\\0& \frac{1}{4}\ln(\frac{1}{4}(D^2-1))
    \end{bmatrix}
    &=g\left(\begin{bmatrix}
        0&D\\-D& 0
    \end{bmatrix}\right)\\&=S^{-1}g(J\sigma)S.
    \end{split}
\end{align}
These results motivate defining the \textit{skewed correlation matrix} $\alpha$,
    \begin{equation}\label{skewCorrMat}
        \alpha = J\sigma = \begin{bmatrix}
            V_{off}^T&P\\-X&-V_{off}
        \end{bmatrix}.
    \end{equation}
    We can now write the logarithm of the density operator in a much more concise form,
    \begin{equation}\label{LogOperator}
    \ln(\hat{\rho}) = -\Trace_{C^2}[\hat{\vectorbold{V}}^Tf(\alpha)J\hat{\vectorbold{V}}+g(\alpha)\hat{I}].
\end{equation}
The nonconstant term is often referred to as the \textit{modular Hamiltonian} $\hat{\mathcal{H}}$ in the literature \cite{Casini_2008},
\begin{equation}\label{modHam}
   \hat{\mathcal{H}} =  \Trace_{C^2}\left[\hat{\vectorbold{V}}^Tf(\alpha)J\hat{\vectorbold{V}}\right].
\end{equation}
Note that, as can be checked from the left hand side of Equation \eqref{SymLogRho}, the operator $f(\alpha)J$ is positive definite

\subsection{Average of the Vacuum Modular Hamiltonian}\label{modhamExp}
Suppose we have a centered Gaussian state $\rho$ with skewed correlation matrix $\alpha$ along with a `vacuum' centered Gaussian state $\rho_0$ with skewed correlation matrix $\alpha_0$ and modular Hamiltonian $\hat{\mathcal{H}}$. We want to compute the quantity $K$ which Casini defined as,
\begin{equation}
    K= \langle\hat{\mathcal{H}}\rangle_{\hat\rho}-\langle\hat{\mathcal{H}}\rangle_{\hat{\rho}_{0}}.
\end{equation}

We use Equation \eqref{modHam} to find the expectation value of the modular Hamiltonian.
To do this, we need to note that $f(\alpha_0)J$ is a symmetric matrix. This is most obvious from comparing the left hand side of \ref{SymLogRho}, with its final right hand side,
\begin{align}\label{modHamExpVal}
\begin{split}
\Trace_{C^2}\left[\langle\hat{\vectorbold{V}}^Tf(\alpha_0)J\hat{\vectorbold{V}}\rangle_{\hat\rho}\right]&=\sum_{jk}(f(\alpha_0)J)_{jk}\langle\hat{V}_j\hat{V}_k\rangle_{\hat\rho}
\\
&=\frac{1}{2}\sum_{jk}(f(\alpha_0)J)_{jk}\langle\{\hat{V}_j\hat{V}_k\}\rangle_{\hat\rho}
\\ &=\frac{1}{2}\Trace_{C^2}\left[f(\alpha_0)\alpha\right]
\end{split}
\end{align}
We can now write Casini's $K$ as,
\begin{equation}\label{CasiniK}
    K = \frac{1}{2}\Trace_{C^2}\left[f(\alpha_0)(\alpha-\alpha_0)\right].
\end{equation}

\subsection{Entropy }\label{GaussEntropy}
Consider a centered Gaussian state $\rho$ with skewed correlation matrix $\alpha$. We want to compute the entropy $S(\rho)=-\langle\ln(\rho)\rangle_{\hat\rho}$. We use our expression for the logarithm of a Gaussian state in Equation \eqref{LogOperator},
\begin{equation}
    S(\rho) =\Trace_{C^2}\left[\langle\hat{\vectorbold{V}}^Tf(\alpha)J\hat{\vectorbold{V}}\rangle_{\hat\rho}+g(\alpha) \langle\hat{I}\rangle_{\hat\rho}\right].
\end{equation}
Clearly, $\langle\hat{I}\rangle_{\hat\rho}=1$. We also know how to take the expectation value of the modular Hamiltonian due to Equation \eqref{modHamExpVal}.
This lets us rewrite the entropy entirely in terms of the skewed correlation matrices as,
\begin{equation}
  S(\rho) = \Trace_{C^2}\left[h(\alpha)\right],
\end{equation}
where we have defined a function $h(z)=\frac z2f(z)+g(z)=\frac{z}{2}\arccot(z)+\frac{1}{4}\ln(-\frac{1}{4}(z^2+1))$. Note that the eigenvalues of $\alpha$ will be imaginary with absolute value larger than one due to the results of Section \ref{Takagi}, making $h(\alpha)$ well-defined.

\subsection{Mutual Information}\label{MI}
Consider a centered Gaussian state $\rho$ on a Hilbert space whose configuration space can be decomposed as $C = C_1\oplus C_2$. As described in Section \ref{block}, we can decompose the skewed correlation matrix $\alpha$ in block matrix form. If we order the basis $C$ such that the $\Omega_1$ vectors come before the $\Omega_2$ vectors, we can write $\alpha$ in block matrix form,
\begin{equation} 
    \alpha = \begin{bmatrix}
        \alpha_{11}& \alpha_{12}\\ \alpha_{21}& \alpha_{22}
    \end{bmatrix}.
\end{equation}
We can now write the skewed correlation matrix $\alpha_{NI}$ of the non-interacting state $\hat\rho_1\otimes\hat\rho_2$,
\begin{equation} 
    \alpha_{NI} = \begin{bmatrix}
        \alpha_{11}& 0\\ 0& \alpha_{22}
    \end{bmatrix}.
\end{equation}
Now, we use the expression for entropy found in Section \ref{GaussEntropy} to write the mutual information between the two systems,
\begin{align}\label{mutualInfoForm}
    \begin{split}
        I(1,2) &= S(\hat\rho_1)+S(\hat\rho_2)-S(\hat\rho)\\&= \Trace_{C^2}\left[h(\alpha_{11})+h(\alpha_{11})-h(\alpha)\right]\\&=\Trace_{C^2}\left[h(\alpha_{NI})-h(\alpha)\right].
    \end{split}
\end{align}

\subsection{Perturbing a  Correlation Matrix Function}\label{generalexpan}
Suppose that we have two skewed correlation matrices $\alpha_0$ and $\alpha_1$. Furthermore, assume the spectral decomposition of $\alpha_0$ is known. Define a matrix valued function $\alpha(t) = t\alpha_1 +(1-t)\alpha_0$. Equivalently, $\alpha(t) = \alpha_0+t\Delta \alpha$ with $\Delta \alpha = \alpha_1-\alpha_0$. For all $t\in [0,1]$, we can show that $-J\alpha(t)+iJ$ is positive definite as shown here:
\begin{equation}
   -J( t\alpha_1+(1-t)\alpha_0) +iJ = t(J\alpha_1 +iJ) + (1-t)(J\alpha_0 +iJ) \geq 0.
\end{equation}
In other words, for all $t\in [0,1]$, $-J\alpha(t)$ is a valid correlation matrix of some Gaussian state. Therefore, we can apply the results of Appendix \ref{Takagi} to the spectrum of $\alpha(t)$. Namely, $\alpha(t)$ is diagonalizable and its spectrum lies in the following set:
\begin{equation}\label{PossibleSpec}
    \mathrm{Spec}(\alpha(t))\subseteq \{z \in \mathbb{C}| z = \pm \lambda i, \lambda > 1\}.
\end{equation}
Given some some matrix function $p$, define $\bar{p}(t)=\Trace_{C^2}[p(\alpha(t))]$ where $\Trace_{C^2}$ is a trace over two copies of the configuration space $C$.\footnote{There is no well-defined notion of trace for all of $C^2$ for a quantum field theory. We once again proceed formally and handle this problem in Section \ref{smooth} under certain assumptions.} The goal of this section is to find a series expansion $\bar{p}(t)=\sum_{n=0}P_n t^n$. The standard  chain rule does not generalize to matrix functions and the non-commutativity of operators makes using a series expansion of $f$ difficult. Instead we write $\bar{p}(t)$ as the following contour integral analogously to Equation \eqref{FuncContour}:
\begin{equation}\label{contPos}
   \bar{p}(t)=\frac{1}{2\pi i}\oint_\xi \diff z \text{ } p(z)\Trace_{C^2}[(z-\alpha(t))^{-1}].
\end{equation}
Here $\xi$ is any contour enclosing the portion of the imaginary line with absolute value greater than one.\footnote{This set is unbounded meaning that realistically integrating over $\xi$ should be viewed as the limit of integrals over bounded contours. We will overlook such details here.} In general, the derivative of a matrix inverse is $\frac{d}{dt}(B(t)^{-1})=$   
$-B(t)^{-1}B(t)$   $B(t)^{-1}$. Using this formula and the product rule for matrices,\footnote{$\frac{d}{dt}(B_1(t)B_2(t))=B_1'(t)B_2(t)+B_1(t)B_2'(t)$} we can derive the derivatives of $\bar{p}(t)$,
\begin{equation}
    \bar{p}^{(n)}(t)=\frac{n!}{2\pi i}\oint_\xi \diff z \text{ } p(z) \Trace_{C^2}\left[(z-\alpha(t))^{-1} \left[\Delta\alpha(z-\alpha(t))^{-1}\right]^n\right].
\end{equation}
This formula can be proved by induction. Put in terms of $P_n$, this formula becomes
\begin{equation}
    P_n=\frac{1}{2\pi i}\oint_\xi \diff z \text{ } p(z)\Trace_{C^2}\left[(z-\alpha_0)^{-1} \left[\Delta\alpha(z-\alpha_0)^{-1}\right]^n\right].
\end{equation}
We substitute $t=1$ and sum over all $n$ to get
\begin{equation}\label{perturbed}
    \bar p(1)=\Trace_{C^2}[p(\alpha_1)] = \frac{1}{2\pi i}\sum_{n=0}^\infty\oint_\xi \diff z \text{ } p(z)\Trace_{C^2}\left[(z-\alpha_0)^{-1} \left[\Delta\alpha(z-\alpha_0)^{-1}\right]^n\right].
\end{equation}
If we look at Equation \eqref{perturbed}, we see it has the form of a geometric series. Summing this series returns us to Equation \eqref{contPos} with $t=1$. This serves as a consistency check but also suggests this formula may be much more generally applicable to perturbative problems than suggested by its derivation here. However, Equation \eqref{perturbed} turns out not to be the most convenient form for the expansion. To find a better one, we take a closer look at each $P_n$. Firstly, if $n=0$, then
\begin{equation}
    P_0 = \frac{1}{2\pi i}\oint_\xi \diff z \text{ } p(z)\Trace_{C^2}\left[(z-\alpha_0)^{-1} \right]= \Trace_{C^2}[p(\alpha_0)].
\end{equation}
For $n\geq 1$, we can use the cyclic invariance of the trace to rewrite $P_n$. We then average over all of the ways to write $P_n$ for different orderings and get
\begin{equation}
    P_n=\frac{1}{2\pi n i }\oint_\xi \diff z \text{ }p(z)\sum_{j=1}^n\Trace_{C^2}\left[\left[\Delta\alpha(z-\alpha_0)^{-1}\right]^{j-1}\Delta\alpha(z-\alpha_0)^{-2}\left[\Delta\alpha(z-\alpha_0)^{-1}\right]^{n-j}\right].
\end{equation}
But now observe that the sum inside the contour integral is the derivative of 
$-\Trace_{C^2}\left[\left[\Delta\alpha(z-\alpha_0)^{-1}\right]^{n}\right]$ with respect to $z$. Therefore, we can use integration by parts\footnote{For a closed contour $\xi$ and holomorphic functions $f$ and $g$, we have $\oint_\xi \diff z \text{ } f(z)g'(z)=-\oint_\xi \diff z \text{ } f'(z)g(z)$. This can be proven by explicitly parameterizing the contour.  } to rewrite this integral as 
\begin{equation}
    P_n=\frac{1}{2\pi n i }\oint_\xi \diff z \text{ }p'(z)\Trace_{C^2}\left[\left[\Delta\alpha(z-\alpha_0)^{-1}\right]^{n}\right].
\end{equation}
For $n=1$, this gives us a particularly nice result:
\begin{equation}
    P_1=\frac{1}{2\pi i }\oint_\xi \diff z \text{ }p'(z)\Trace_{C^2}\left[\Delta\alpha(z-\alpha_0)^{-1}\right] = \Trace_{C^2}\left[p'(\alpha_0)\Delta\alpha\right].
\end{equation}
We can therefore write a general series expansion for $\Trace_{C^2}[p(\alpha_1)]$ as
\begin{equation}\label{perturbedNice}
    \Trace_{C^2}[p(\alpha_1)]=\Trace_{C^2}[p(\alpha_0)]+\Trace_{C^2}\left[p'(\alpha_0)\Delta\alpha\right]+\sum_{n=2}^\infty\frac{1}{2\pi n i }\oint_\xi \diff z \text{ }p'(z)\Trace_{C^2}\left[\left[(z-\alpha_0)^{-1}\Delta\alpha\right]^{n}\right].
\end{equation}

In summary, this formula describes a perturbative series for the entropy of $\alpha_1$ in terms of the resolvent of $\alpha_0$ and the difference of skewed correlation matrices $\Delta\alpha=\alpha_1-\alpha_0$. 

\subsection{Diagonalizing the Skewed Correlation Matrix via the Modular Hamiltonian}\label{sec:modHamLatt}
Recall the form of the modular Hamiltonian in terms of the skewed correlation matrix as 
\begin{equation}
     \hat{\mathcal{H}} =  \hat{\vectorbold{V}}^Tf(\alpha)J\hat{\vectorbold{V}},
\end{equation}
where the matrix $f(\alpha)J$ is positive definite. Note that we have dropped the unnecessary configuration space trace, as the operator is already a scalar-valued operator. Suppose that, for some reason, the spectral decomposition of the operator appearing in the modular Hamiltonian, $f(\alpha)$, is easier to compute than $\alpha$. How do we compute the spectral decomposition of $\alpha$ in terms of the spectral decomposition of $f(\alpha)$? 

Denote by $\langle \cdot,\cdot\rangle$ the following inner product on $C^2$,
\begin{equation}
    \langle v,w\rangle = v^\dagger Jf(\alpha) w.
\end{equation}
This is a sensible inner product because $f(\alpha)J $ is positive definite and $J^\dagger=J^{-1}=-J$, making $ Jf(\alpha)$ positive definite as well. $f(\alpha)$ is skew-Hermitian with respect to this inner product, which follows from
\begin{equation}
    f(\alpha)^\dagger (Jf(\alpha)) = -(Jf(\alpha))^\dagger f(\alpha) = -(Jf(\alpha))f(\alpha),
\end{equation}
where we make use of the fact that $Jf(\alpha)$ is symmetric.
This implies the eigenvalues of $f(\alpha)$ are imaginary. Granted, we knew this already because $\alpha$ has imaginary eigenvalues and $f$ is an odd function, but it is a good consistency check nonetheless. Now, consider the eigenvalue problem for $f(\alpha)$:
\begin{equation}\label{ModHamEigs}
    f(\alpha)\phi_k = i\lambda_k\phi_k,
\end{equation} 
where we label the eigenvalues with the index $k$.
Since $f(\alpha)$ is skew-Hermitian, its eigenvectors are orthogonal with respect to the $Jf(\alpha)$ inner product, a fact which can be rewritten in terms of the standard inner product in $C^2$ as
\begin{equation}
 v_{k'}^\dagger Jf(\alpha) v_{k} =\delta_{k,k'}.
\end{equation}
But, using the fact that $v_{k}$ is an eigenvector of $f(\alpha)$, we can remove $f(\alpha)$ from the orthogonality relation:
\begin{equation}\label{Normalization}
        -i\lambda_k (Jv_{k'})^\dagger v_{k} =\delta_{k,k'}.
\end{equation}
This orthogonality relation allows us transform into the $\alpha$-eigenbasis. Suppose that we separate the eigenvectors $v_k$ into its two component vectors on $C$,
\begin{equation}
  v_k = \begin{bmatrix}
        v_{k,1}\\ v_{k,2}
    \end{bmatrix}.
\end{equation}
Then, given any matrix $A$ in $C^2$, we can write the entries in the $f(\alpha)$ eigenbasis as
\begin{equation}
    A_{k,k'} = \begin{bmatrix}
         v_{k',2}^\dagger&-v_{k',1}^\dagger  \end{bmatrix}A\begin{bmatrix}
        v_{k,1}\\  v_{k,2}
    \end{bmatrix}.
\end{equation}
In particular, we can write the resolvent of $\alpha$ in the $f(\alpha)$ eigenbasis as
\begin{equation}\label{DiagonalKernel}
    ((z-\alpha)^{-1})_{k,k'}= \frac{\delta_{k,k'}}{z+i\coth(\lambda_k)}.
\end{equation}

\section{Mathematical Background}
\label{app:math}

\subsection{Symplectic Matrices}\label{sec:symp}
Let $C$ be a $n$-dimensional complex vector space. We define the following operator in $C^2=C\oplus C$,
\begin{equation}\label{eq:J}
    J =\begin{bmatrix}
        0& I\\-I&0\\
    \end{bmatrix},
\end{equation}
where $I$ is the identity matrix on $C$. Observe that $J^2 = -1$. We say an operator is \textit{symplectic} if it preserves the bilinear form $v^\dagger J w$. Equivalently, a symplectic operator is an operator $S$ satisfying the equation $S^\dagger J S = J$. The real symplectic matrices of dimension $2n$ form a Lie group denoted $SP(2n,\mathbb{R})$. The Lie algebra of this group is denoted by $\mathfrak{sp}(2n,\mathbb{R})$ and consists of real matrices $A$ satisfying $JA+A^T J=0$, the symplectic analogues of skew-symmetric matrices. Mathematicians call these matrices \textit{Hamiltonian}. This nomenclature would be confusing due to the Hamiltonian energy operator however. So we will avoid it and simply say `in $\mathfrak{sp}(2n,\mathbb{R})$' instead.

To see the relevance of symplectic matrices to quantum mechanics, suppose we are working in a quantum system with configuration space $C$. We define the operator valued vector $\hat{\vectorbold{V}}=(\hat{\boldsymbol{\phi}},\hat{\boldsymbol{\pi}})$. For finite dimensions, this is a vector in $C^2$. The canonical commutation relation can be written as the following equivalence between operator valued matrices in $C^2$:
\begin{equation}\label{uncertainty}
    \frac{1}{i}[\hat{V}_j,\hat{V}_k] = J_{jk}.
\footnote{We set $\hbar = 1$ everywhere in this paper.}\end{equation}

This relationship defines a bilinear form in the configuration space. A Bogoliubov transformation is a linear transformation in the configuration space that preserves this commutation relation. In other words, a Bogoliubov transformation is a linear canonical transformation in the classical variables.

\subsection{The Symplectic Takagi Factorization}\label{Takagi}
Let $C$ be a $n$-dimensional complex vector space and suppose that we have a real positive definite matrix $A$ on $C^2$. Then, we can define an inner product $\langle v,w\rangle = v^\dagger A w$ on $C^2$. Now, consider the matrix $JA$. We can see that this matrix is skew-Hermitian with respect to this inner product, which can be seen here: \begin{equation}
    \langle JAv,w \rangle = -v^\dagger A J A w =  -\langle v,JAw \rangle.
\end{equation}
$JA$ is also clearly real. Therefore, by the spectral theorem for skew Hermitian matrices, $JA$ is diagonalizable with eigenvalues $\pm i\lambda_1,\dots,\pm i\lambda_n$. Given the matrix $D_0=\rm{diag}(\lambda_1,\dots,\lambda_n)$, we define a matrix $W$.
\begin{equation}
    W = \begin{bmatrix}
        0& D_0\\-D_0&0\\
    \end{bmatrix}
\end{equation}
$W$ and $JA$ have the same eigenvalues and are both diagonalizable. So, they are similar. Furthermore, direct calculation shows that they are both in $\mathfrak{sp}(2n,\mathbb{R})$. Therefore they are symplectically similar, by \cite{HORN199543}. That is to say, there exists a real symplectic matrix $S$ such that  
\begin{equation}\label{similiar}
    W=S^{-1}JAS.
\end{equation}
Now from the definition of symplectic, we have $S^T=-JS^{-1}J$. This allows us to find a symplectic matrix $S$ such that $S^TAS$ is diagonal, as shown here:
\begin{equation}
    S^TAS = -JS^{-1}JAS = -JW = \begin{bmatrix}
        D_0& 0\\0&D_0\\
    \end{bmatrix}.
\end{equation}
We will call this diagonal matrix $D$. This is the called symplectic Takagi Factorization of $A$.

In this paper, we are particularly interested in positive definite matrices $A$ such that $A+iJ$ is also positive definite. In this case, $S^T(A+iJ)S=D+iJ$ is also positive definite. This is true if and only if each $\lambda_i > 1$.

\subsection{Block Matrices of a Product}\label{Blockpath}
Let $C$ be a vector space decomposed into a direct sum of closed subspaces $C = \bigoplus_{i=1}^n C_i$. Suppose we have a sequence of matrices $A^1,A^2,\dots,A^m$ in $C$ and we know their blocks $A^k_{i_1,i_2}$. We want to find the block matrix from $C_i$ to $C_j$ of the product $A^1\dots A^m$? To do this, we consider a sum over all paths of length $m$ from $C_i$ to $C_j$. Such paths can represented schematically as
\begin{equation}
    i\to i_1 \to \dots\to i_{m-1}\to j,
\end{equation}
where $i,i_1,\dots,i_{m-1},j$ label subspaces.
The block matrix of the product can now easily be written as
\begin{equation}
    (A^1\dots A^m)_{ij} = \sum_{i_1,\dots, i_{m-1}} A^1_{i,i_1}\dots A^m_{i_{m-1},j}.
\end{equation}

\subsection{Writing a Matrix Function as a Contour Integral}\label{FunConSect}
Let $C$ be a finite dimensional vector space. Given a diagonalizable matrix $A$ on $C$ with eigenvalues $\lambda_k$ and eigenvectors $v_k$, we want to find the matrix function $f(A)$ for an analytic function $f$. The standard approach is to use the diagonalization $A=QDQ^{-1}$ and simply write $f(A)=Qf(D)Q^{-1}$, where $f(D)$ is diagonal with $(f(D))_{ii}=f(\lambda_i)$. $f(A)$ can be written as a sum,
\begin{equation}
    f(A)=\sum_k v_k w_k^\dagger f(\lambda_k),
\end{equation}
where $w_k$ are the columns of $(Q^{-1})^T$. However, this approach does not lend itself easily to perturbations, as perturbing the matrix $A$ changes both the eigenvalues and the eigenvectors. We will use an alternative formula for $f(A)$ assuming $f$ is holomorphic on some open set containing the eigenvalues of $A$. We write $f(A)$ using a contour integral over $\xi$, which is any contour enclosing all eigenvalues of $A$,
\begin{equation}\label{FuncContour}
    f(A) = \frac{1}{2\pi i }\oint_\xi \diff z \text{ } f(z) (z-A)^{-1}.
\end{equation}
This equation can be proven by decomposing $(z-A)^{-1}$ into index form and using the residue theorem,
\begin{equation}
\frac{1}{2\pi i }\oint_\xi \diff z \text{ } f(z) (z-A)^{-1} = \sum_k v_k w_k^\dagger \left[\frac{1}{2\pi i }\oint_\xi \diff z \frac{f(z)}{z-\lambda_k}\right]=\sum_k v_k w_k^\dagger f(\lambda_k)=f(A).\end{equation}

\subsection{Spherical Harmonics in Higher Dimensions}\label{highSim}
Consider the surface of the unit sphere in $d$ dimensions which we label $\partial \Omega$. We can parameterize $\partial \Omega$ using hyperspherical coordinates.
\begin{align}
\begin{split}
  x_1 &=  \cos(\theta_1) \\
  x_2 &= \sin(\theta_1) \cos(\theta_2) \\
  x_3 &= \sin(\theta_1) \sin(\theta_2) \cos(\theta_3) \\
      &\quad\quad\quad\quad\quad\vdots\\
  x_{d-1} &= \sin(\theta_1) \cdots \sin(\theta_{d-2}) \cos(\theta_{d-1}) \\
  x_d     &= \sin(\theta_1) \cdots \sin(\theta_{d-2}) \sin(\theta_{d-1}) .
  \end{split}
\end{align}
Necessary in this construction is an ordering of the axes $x_1,x_2,\dots$ until we reach the final two axes $x_{d-1}$ and $x_{d}$ to which we place equal importance. From this ordering, we can naturally construct $d$-dimensional matrix representations of the rotation groups $SO(d),SO(d-1),\dots ,SO(2)$. We represent $SO(d)$ by all rotations, $SO(d-1)$ by the rotations that keep the $x_1$ axis fixed,  $SO(d-2)$ by the rotations that keep the $x_1$ and $x_2$ axes fixed and so on.

Consider the set of square integrable functions $S_{\partial \Omega}$ on $\partial \Omega$. Given each rotation group $G$ in the sequence above with a $d$ dimensional matrix representation $Q$, we can naturally define a corresponding representation on $S_{\partial \Omega}$ by rotation of the input.\footnote{That is, we represent $g$ with the group action $R_g[\phi](\vectorbold{x})=\phi(Q_g^T\vectorbold{x})$.} 
The quadratic Casimir invariants of these representations form a complete set of commuting Hermitian matrices\footnote{A set of commuting Hermitian matrices is `complete' if there is they have at most one simultaneous eigenvector for any combination of their eigenvalues. }. Their simultaneous eigenvectors are called the spherical harmonics and form a symmetry adapted basis of $S_{\partial \Omega}$ for each representation of $SO(n)$. The spherical harmonics are labeled by the indices $\ell, m,\eta_1, \eta_2,\dots, \eta_{d-3}$. Spherical harmonics of constant $\ell$ form a basis for an irreducible representation of $SO(d)$, harmonics of constant $\ell$ and $m$ label an irreducible representation $SO(d-1)$, and so on. In this work only the first two indices are important to isolate and we will compress the remaining indices into a single $\eta$, labeling the spherical harmonics as $Y_{\ell}^{m,\eta}$.

For $d=2$, $\ell$ can be any integer and the remaining indices are superfluous. We will define them to always be zero. For $d=3$, $\ell$ and $m$ are integers satisfying $\ell\geq |m|$, while $\eta$ is superfluous and will also be defined to always be zero. For $d\geq 4$, $\ell$ and $m$ are non-negative integers satisfying $\ell\geq m$.

The spherical harmonics are normalized such that $\int_{\partial \Omega}\diff S \text{ } |Y_{\ell}^{m,\eta}|^2 =1$ and they form an orthonormal basis for $S_{\partial \Omega}$.
In the case of $m=\eta=0$, a spherical harmonic has a relatively simple form in spherical coordinates. For $d\geq 3$, they are proportional to Gegenbauer polynomials,
\begin{equation}\label{harmonicD>2}
    Y_\ell^{0,0}(\theta_1,\dots,\theta_{d-1}) = B_{d,\ell}C_\ell^{\frac{d}{2}-1}(\cos(\theta_1)).
\end{equation}
Here a Gegenbauer polynomial $C_\ell^{a}$ is the degree $\ell$ classical orthogonal polynomial on the interval $[-1,1]$ such that $\frac{1}{(1-2xt+t^2)^a}=\sum_{\ell=0}^\infty C_{\ell}^\alpha(x)t^\ell$.
We also define a normalization constant, 
\begin{equation}\label{B}
B_{d,\ell}=\pi^{\frac{1-d}{2}}2^{-\frac{d}{2}}\Gamma(d-2)\sqrt{\frac{\Gamma(\ell+1)(d-2+2\ell)}{\Gamma(d-2+\ell)}}.\end{equation} 
For $d=2$, they are complex exponentials,
\begin{equation}\label{harmonicD=2}
    Y_\ell(\theta_1,\dots,\theta_{d-1}) = \frac{1}{\sqrt{2\pi}} e^{i\ell \theta_1}.
\end{equation}
However, they are still related to the Gegenbauer polynomials through a limit,
\begin{equation}\label{lim2D}
   \lim_{d\to 2^+}B_{d,\ell}C_\ell^{\frac{d}{2}-1}(\cos(\theta_1)) =\begin{cases}
       \frac{1}{\sqrt{2}}(Y_\ell+Y_{-\ell}) & \ell \neq 0 \\Y_0 & \ell = 0
   \end{cases}.
\end{equation}
We point out the the expression in the limit on the left hand side is well-defined for all real numbers $d > 2$, not just integers.
We define $\mathcal{N}(d,\ell)$ as dimension of the representation of $SO(d)$ composed of $d$ dimensional spherical harmonics of rank $\ell$. Note that this is both the number of $d$-dimensional spherical harmonics with first index $\ell$ and the number of $d+1$ dimensional harmonics with first index $\ell_0$ and second index $\ell$. There exists a simple formula for $\mathcal{N}(d,\ell)$ \cite{frye2012spherical}, using the binomial coefficient ${n \choose k}$,\footnote{We take ${n \choose k}=0$ if $n<k$.} 
\begin{equation}\label{numHarm}
    \mathcal{N}(d,\ell)= \begin{cases}

        {d+\ell-1 \choose d-1}-{d+\ell-3 \choose d-1} & d\geq 3\\ 1 & d=2  \end{cases}.
\end{equation} 
We highlight a useful property of the spherical harmonics. Any function on the boundary of the sphere has a unique harmonic extension to the sphere's interior, due to the Poisson kernel formula \cite{Axler_Bourdon_Ramey_2001}. The extension of a spherical harmonic $Y_\ell^{m,\eta}$ is the \textit{solid harmonic} $R_\ell^{m,\eta}$,
\begin{equation}
    R_\ell^{m,\eta}(r,\theta_1,\dots,\theta_{d-1})=r^{|\ell|} Y_\ell^{m,\eta}(\theta_1,\dots,\theta_{d-1}).
\end{equation}
The solid harmonics $R_\ell^{m,\eta}$ are special because they are homogeneous $d$-variate polynomials of degree $\ell$. Note that the function $f(r)=r^2=\sum_i x_i^2$ is also a homogeneous polynomial. By combining these polynomials we can construct the following set of homogeneous polynomials using two indices satisfying $j\leq \lfloor\frac{n}{2}\rfloor$,
\begin{equation}\label{homBasis}
     G_{n,j}^{m,\eta}(\vectorbold{x})=|\vectorbold{x}|^{2j}R_{n-2j}^{m,\eta}(\vectorbold{x}).
\end{equation}
The exception is for $d=2$, where we must be careful to account for the fact that the index of a solid harmonic can be negative. The set of homogeneous polynomials are in this case,
\begin{equation}
      G_{n,j}(\vectorbold{x})= \begin{cases}
|\vectorbold{x}|^{2j}R_{n-2j}^{0,0}(\vectorbold{x})&n \geq 0 \\
|\vectorbold{x}|^{2j}R_{-n+2j}^{0,0}(\vectorbold{x})&n < 0
      \end{cases}.
\end{equation}
On one hand, these polynomials are all linearly independent because they are in different wave mode subspaces. On the other hand, we can compute the total number of these homogeneous polynomials with a telescoping sum via Equation \eqref{numHarm}, giving us\footnote{For $d=2$, we can directly calculate the number of $G_{\ell,j}$ of a certain degree and we get $\ell+1$, which also produces the right hand side ${2+\ell-1 \choose 2-1}$.}
\begin{equation}
    \sum_{j=0}^{\lfloor\frac{n}{2}\rfloor}\mathcal{N}(d,n-2j)= {d+n-1 \choose d-1}.
\end{equation}
This is exactly the dimension of the space of $d$-variate homogeneous polynomials of degree $\ell$ by a combinatorial stars and bars argument. Therefore, the functions $G_{\ell,j}^{m,\eta}$ form a basis for this space. Since any polynomial can be written as a sum of homogeneous polynomials, these functions can also serve as a basis for all multivariate polynomials if we allow $\ell$ to vary.
\subsection{Accounting for Spherical and Cylindrical Symmetry on a Sphere}\label{spher}
In this work, we deal with quantum field theories on the unit sphere, denoted by $\Omega$ in arbitrary dimensions. In this case, it will be useful to decompose fields into spherical wave modes. We recall from the main body that the configuration space of fields is the set of tempered distributions on the sphere. Given any tempered distribution on the sphere, written in spherical coordinates $\phi(r,\theta_1,\dots,\theta_{d-1})$,  we can (formally) decompose it in terms of spherical harmonics\footnote{Any Schwartz function can be decomposed into wave modes by decomposing into the orthonormal basis of spherical harmonics on each spherical shell. This decomposes the test space into closed subspaces and we can decompose the tempered distributions by duality.},
\begin{equation}\label{harmdec}
    \phi(r,\theta_1,\dots,\theta_{d-1})=\sum_{\ell,m,\eta}\psi_{\ell,m,k}(r)Y_\ell^{m,\eta}(\theta_1,\dots,\theta_{d-1}),
\end{equation}
where $\psi_{\ell,m,k}(r)$ is a radial component function.
This allows us to decompose the configuration space $C_\Omega$ into the wave mode subspaces $C_{\ell,m,\eta}$ consisting of the fields of the form $\psi(r)Y_\ell^{m,\eta}(\theta_1,\dots,\theta_{d-1})$ for some one dimensional tempered distribution $\psi(r)$,
\begin{equation}
    C_\Omega = \bigoplus_{\ell.m,\eta} C_{\ell,m,\eta}.
\end{equation}

Recall the definition of $C_\Omega^2$, the space of tempered-distribution-valued 2-vectors. The decomposition of $C_\Omega$ gives a decomposition of $C_\Omega^2$ into the wave mode subspaces $C^2_{\ell,m,\eta} =  C_{\ell,m,\eta}\bigoplus  C_{\ell,m,\eta}$ through the \textit{spherical wave mode decomposition},
\begin{equation}
    C_\Omega^2 = \bigoplus_{\ell.m,\eta} C_{\ell,m,\eta}^2.
\end{equation}

Thanks to this decomposition, we can naturally the generalize the notions of block matrices given in Section \ref{block}. Given some operator $A$ on the configuration space with $2\times 2$ matrix valued kernel $K(\vectorbold{x},\vectorbold{y})$, we can define a block matrix from $C_{\ell,m,\eta}^2$ to $C_{\ell',m',\eta'}^2$ with respect to the spherical wave mode decomposition. We can write this block matrix in terms of a $2\times 2$ matrix-valued \textit{radial kernel} $K_{\ell,m,\eta,\ell',m',\eta'}$ operating on the radial component of the vector in $C_{\ell,m,\eta}^2$. This block matrix can be computed by integrating twice over the unit sphere,
\begin{equation}
    K_{\ell,m,\eta,\ell',m',\eta'}(s,r)=\int_{|\vectorbold{x}|=1} \diff^d \vectorbold{x}\int_{|\vectorbold{y}|=1} \diff^d \vectorbold{y} \text{ } K(r\vectorbold{x},s\vectorbold{y})Y_{\ell}^{m,\eta}(\vectorbold{x})Y_{\ell'}^{m',\eta'}(\vectorbold{y})^*.
\end{equation}

Given any rotation group $G$ with a $d$-dimensional orthogonal representation $Q$, we can define a corresponding representation $C_\Omega^2$ by rotation of the input, similar to Section \ref{highSim}. In this work, we explores cases where a matrix $A$ is invariant with respect to some representation $R$. Explicitly this means that $AR_g=R_gA$ for every $g\in G$. In this work we are interested in the cases when $G=SO(d)$ and $Q$ is its natural representation and $G=SO(d-1)$ and $Q$ is the representation of rotation around the $x_1$ axis (i.e. the representations labeled by the first and the first two indices of the spherical harmonics respectively).

Due to an important result in representation theory called Schur's Lemma, we can find selection rules for $K_{\ell,m,\eta,\ell',m',\eta'}$ in these two cases.\footnote{For an exposition of Schur's lemma for the case of a finite dimensional vector space, see Section 2 of Stiefel and Fassler \cite{Fassler1992}.}. For $G=SO(d)$, $K_{\ell,m,\eta,\ell',m',\eta'}=0$ unless $\ell=\ell'$, $m=m'$ and $\eta=\eta'$. Furthermore, it is independent of $m$ and $\eta$. For $G=SO(d-1)$, $K_{\ell,m,\eta,\ell',m',\eta'}=0$ unless $m=m'$ and $\eta=\eta'$.\footnote{For $d=2$, this selection rule still holds if one uses our convention that superfluous indices are always zero.} Furthermore, it is independent of $k$. For both of these cases, it will be convenient in the main body to omit the indices which a block is independent of.

\bibliographystyle{JHEP.bst}

\bibliography{misqft.bib}

\end{document}